\documentclass[aps,showpacs,amsmath,amsfonts,amssymb,floatfix,pre]{revtex4}
\usepackage{graphicx}
\usepackage{dcolumn}
\usepackage{bm}
\begin{document}

\title{Entanglement production in coupled chaotic systems : Case of the Kicked
Tops}

\author{Jayendra N. Bandyopadhyay}
\email{jayendra@prl.ernet.in}  
\author{Arul Lakshminarayan 
\footnote{Present Address: Department of Physics, Indian Institute of 
Technology, Madras, Chennai 600036, India.}}
\email{arul@physics.iitm.ac.in} 
\affiliation{Physical Research Laboratory, Navrangpura, Ahmedabad 380009, 
India.}

\date{\today}

\begin{abstract}

Entanglement production in coupled chaotic systems is studied with the
help of kicked tops. Deriving the correct classical map, we have
used the reduced Husimi function, the Husimi function of the reduced density 
matrix, to visualize the
possible behaviors of a wavepacket. We have studied a phase space
based measure of the complexity of a state and used random matrix
theory (RMT) to model the strongly chaotic cases. Extensive numerical
studies have been done for the entanglement production in coupled
kicked tops corresponding to different underlying classical dynamics
and different coupling strengths. An approximate formula, based on
RMT, is derived for the entanglement production in coupled strongly
chaotic systems. This formula, applicable for arbitrary coupling
strengths and also valid for long time, complements and
extends significantly recent perturbation theories for strongly
chaotic weakly coupled systems.

\end{abstract}


\maketitle

\section{Introduction}

A quantum mechanical system, which consists of at least two
interacting subsystems, has an unique property called `entanglement'
\cite{schro}. This property is unique in the sense that even if we
know the exact state of the system, it is in general not possible to
assign any pure state to the subsystems.  Entanglement is a
nonclassical correlation among the subsystems which exists even
between spatially well separated subsystems \cite{epr}. This unique
property of a quantum system has been characterized as a quantum
resource for quantum information theory and quantum computation
\cite{nielsen}.  Moreover, quantum entanglement has also been studied
extensively from the decoherence point of view. It has been argued
that a quantum system in the presence of an ``environment'' can loose
its coherence and behave more like a classical system \cite{zurek}. 

A quantum computer is a collection of many interacting particles. 
Such a many-particle structure may be prone to problems of 
decoherence and chaos. Decoherence can create some errors in the 
operation of a quantum computer, however, these errors, in principle, can be
removed by quantum error correcting codes \cite{nielsen}. 
On the other hand, the problem due to chaos has recently attracted some 
attention. It has been shown that residual, uncontrolled interaction between 
the particles might induce quantum chaos in the quantum computer if the
interaction strength crosses certain critical limits and consequently,
it may destroy the operational condition of the quantum computer
\cite{shepe1}. Besides quantum chaos can also emerge during the
implementation of some quantum algorithms \cite{braun}. Obviously, a
quantum algorithm which simulates a quantum chaotic system is by
definition a unitary operation showing quantum chaos
\cite{shepe2}. However, it has been shown that well known algorithms,
such as Grover's search algorithm and the quantum Fourier transform
algorithm give rise to some unusual combination of quantum signatures
of chaos and of integrability \cite{braun}. The error due to the
presence of chaos in a quantum computer can also be corrected by error
correcting codes, but the presence of chaos enhances the complexity
and hence much more error correction is needed \cite{silves}.
Therefore, the knowledge of the presence and effects of chaos in a
quantum computer is necessary to implement proper error correcting
codes. Very recently, the behavior of quantum entanglement during the
operation of an efficient algorithm for quantum chaos have been
studied \cite{shepe3}. However, here we are interested at a more basic
level to study the effect of the underlying classical dynamics on
entanglement production.

Recently, several studies have explored this question
\cite{furuya,angelo99,angelo01,sarkar1,arul,our,tanaka,a_lahiri,caves}. The
first one studied the entanglement production in an $N$-atom
Jaynes-Cummings model \cite{furuya} and they found that the
entanglement rate was considerably enhanced if the initial wavepacket
was placed in a chaotic region. They also argued that their results
support an earlier conjecture which predicted that the entanglement
production in a chaotic system, coupled to an environment, would be
more than the regular system \cite{zurekpaz}.  According to that
conjecture, the entanglement production rate would be higher for a
chaotic system coupled to an environment. For the $N$-atom
Jaynes-Cummings model, each atomic subsystem plays the role of an
environment for the other. Later, it has been shown that large
entanglement production rate is not the hallmark of a nonintegrable
system \cite{angelo99}. Even in the integrable $N$-atom
Jaynes-Cummings model some special initial coherent states exhibit
strikingly similar entanglement production as
corresponding to the chaotic case \cite{angelo01}.  

In another paper, the entanglement production rate has been related to
the classical Lyapunov exponents with the help of a coupled kicked
tops model \cite{sarkar1}. They also justified their findings on the
basis of the above mentioned conjecture \cite{zurekpaz}. However, the
classical limit of the coupled kicked tops derived in this rather
well-quoted work is incorrect, in fact it is not even
canonical. However they consider very weakly coupled tops and
therefore their conclusions turn out to be qualitatively valid. In
other work, one of us studied the entanglement in coupled standard
maps and found that entanglement increased with coupling strength, but
after a certain magnitude of coupling strength corresponding to the
emergence of complete chaos, the entanglement saturated
\cite{arul}. Similar saturation of entanglement was also observed for
a time evolving state, which was initially unentangled.  This
saturation value depended on the Hilbert space dimension of the
participating subsystems, and was less than its maximum possible
value. It was also pointed out that in analogy with environment
induced decoherence, the reduced density matrices (RDMs) corresponding
to subsystems of fully chaotic systems, are diagonally
dominant. 

Later, we derived the saturation value of the entanglement
using random matrix theory \cite{our}. Moreover, we presented a
universal distribution of the eigenvalues of the RDMs, and
demonstrated that this distribution is realized in quantized chaotic
systems by using the model of coupled kicked tops. Subsequently, an
analytical explanation for the entanglement production, based on
perturbation theory, has been given for two weakly coupled strongly
chaotic systems \cite{tanaka}. The authors also found that 
increase in the strength of chaos does not enhance the entanglement
production rate for the case of weakly coupled, {\it strongly} chaotic,
subsystems. In a recent work, entropy
production in subsystems has been examined as a dynamical criterion
for quantum chaos \cite{a_lahiri}. It has been observed
that the power spectrum of the entropy production gets progressively
broad banded with a progressive transition from regular
to chaotic systems. More recently, entanglement production has been
investigated in a class of quantum baker's map \cite{caves}. They
also found that, in general, the quantum baker's map is a good
dynamical system to generate entanglement.

Besides these studies of entanglement production and decoherence in
coupled chaotic systems, extensive studies have been done on
decoherence of chaotic systems that are coupled to an
environment. These studies were mainly motivated by the fact that
decoherence induces a transition from quantum to classical-like
behavior and therefore, this decoherent approach can be utilized in a
more straightforward way to restore the correspondence between a
quantum chaotic system and its classical counterpart
\cite{zurekpaz}. Irreversibility is the price of this decoherent model
for the restoration of quantum-classical correspondence in a quantum
system. This irreversibility causes entropy production in the
system. It has been conjectured, as already mentioned, that this
entropy, grows linearly in time with a fixed rate determined by the
Lyapunov exponents. 

This conjecture has been tested for several model open quantum
chaotic systems. It has been shown that the entropy production rate,
as a function of time, in a quartic double well with harmonic driving
coupled to a sea of harmonic oscillators has atleast two distinct
regimes \cite{diana}. For short times this rate is proportional to the
system-environment coupling strength, and for longer times there is a
regime where this rate is determined by the Lyapunov exponent. In
another work, the entropy production in the baker's map and Harper's
map coupled to a diffusive environment is studied \cite{paz}. A regime
was found to exist where the entropy production rate is determined by
the system's dynamical properties like Lyapunov exponents, folding
rates, etc., and moreover, in this regime the entropy production rate
becomes independent of the system-environment coupling
strength. Similar results are also reported in
Refs. \cite{arjenduPRL99,sarkarlahiri}. In other work evidence has
been presented that the decoherence rate (or entropy production rate)
of a quantum system coupled to an environment is governed by a
quantity which is a measure of both the increasingly detailed structure of
the quantum distributions (Wigner function) and classical phase space
distributions \cite{arjendu}.

Very recently, it has been reported that, in
open quantum systems, there exists a universal scaling among the
parameters (effective Planck's constant, measure of the coupling
strength between system and environment, classical Lyapunov exponents)
on which the quantum-classical transition of that system depends
\cite{arjenduPRL03}. In another direction, decoherence has been
discussed in an open system coupled to a nonlinear environment with
finite degrees of freedom \cite{kubotani95}. It was found that even
though the environment is finite dimensional, the strong nonlinearity
of it can destroy the quantum coherence. Hence there is a possibility
to utilize this finite dimensional chaotic system as a model of
environment, instead of infinite dimensional heat bath. The above
possibility has also been discussed in a recent work
\cite{cohenkottos}. Naturally this approach is closely linked to
studies like the present one on the coupled kicked tops.

We have discussed two different approaches in the study of
entanglement production and decoherence in chaotic systems. First
approach was to study the entanglement production and decoherence in
coupled chaotic systems by performing exact numerical calculation or
using some model based on random matrix theory and perturbation
theory. The second approach was mainly based on approximate master
equations. In this paper, following the first approach, we have
studied entanglement production in coupled kicked tops. We have
considered the entanglement production for both chaotic and regular
cases. Besides considering the effect of different kind of classical
dynamics on quantum entanglement, we have also considered the effect
of different coupling strengths on entanglement production. We have
extensively studied a measure of the complexity of the time evolving
state, based on the second moment of the Husimi function of that
state, corresponding to both single and coupled tops. Using RMT, we
have explained the behaviors of this measure for strongly chaotic
cases.  We have then derived an analytical formula for the
entanglement production in coupled strongly chaotic systems using
RMT. This formula is applicable for any coupling strength and it also
valid for sufficiently long time.
    
This paper comprises of six sections. In Sec.\ref{couptop}, we have
discussed the quantum and classical properties of coupled kicked
tops. We have presented the correct classical map of the coupled
kicked tops. We have discussed the initial states used and have
defined the measures of entanglement used here. Finally, we have
concluded this section by discussing a method to visualize the
wavepacket of a coupled system on the phase space of a subsystem. In
Sec.\ref{W2}, we have considered a recently proposed method to measure
the complexity of a quantum state. Using this method, we have defined
a measure which quantifies the fraction of the total number of Planck
cells occupied by the Husimi function of a given state, roughly
speaking the amount of ``phase space'' that is filled by the Husimi.

The Hilbert space dimension is the number of Planck cell's, each of
volume $h^d$, that fit into the total phase space volume. In
one-dimension, $d=1$, then $N=\mbox{Phase Space Area}/h$.  The above mentioned
measure of the complexity of quantum states is also approximately
equal to the fraction of the Hilbert space occupied by the given
state. We have observed for the single top that a typical time
evolving state can occupy half of the total number of the Planck cells,
and this happens only for the strongly chaotic cases. Whereas for a
highly chaotic top coupled strongly to another such top, the above
measure, now for the reduced density matrix of each top, has reached a
value very close to unity. We explain the behavior of this measure,
using random matrix theory (RMT), for the strongly chaotic cases.  For
nonchaotic and mixed cases, the time evolving state occupies lesser
number of Planck cells and is reflected in smaller values of this
measure.
 

In the next section, Sec.\ref{numerics}, we have presented the numerical 
results on the entanglement production. In Sec. \ref{analytical}, we have 
derived an approximate formula, based on RMT, to explain the entanglement
production in coupled strongly chaotic systems. Finally, we summarize
in Sec.\ref{last}.

\section{\label{couptop}Coupled kicked tops}

\subsection{\label{QT}Quantum Top}

The single kicked top is a system, characterized by an angular
momentum vector ${\bf J} = ( J_x, J_y, J_z )$, where these components
obey the usual angular momentum algebra. We set  Planck's constant
to unity. The dynamics of the top is governed by the
Hamiltonian \cite{haake1}:
\begin{equation}
H(t) = p J_y + \frac{k}{2j} J_{z}^{2} \sum_{n=-\infty}^{+\infty} \delta (t-n).
\label{single}
\end{equation} 
\noindent The first term describes free precession of the top around $y-$axis
with angular frequency $p$, and the second term is due to periodic 
$\delta$-function kicks. Each such kick results in a  
torsion about $z-$axis by an angle proportional to $J_z$, and the 
proportionality factor is a dimensionless constant $k/2j$. Now, to study the
entanglement between two tops, we consider the Hamiltonian of the coupled 
kicked tops which can be written, following Ref. \cite{sarkar1}, as :
\begin{eqnarray}
{\cal H}(t) &=& H_1(t) + H_2(t) + H_{12}(t)\\
\mbox{where}~~H_i(t) &\equiv& p_i J_{y_i} + \frac{k_i}{2j} J_{z_i}^{2} 
\sum_{n} \delta (t-n),\\
H_{12}(t) &\equiv& \frac{\epsilon}{j} J_{z_1} J_{z_2} \sum_{n} \delta (t-n),
\label{couple}
\end{eqnarray}
\noindent where $i = 1, 2$. Here $H_i(t)$'s are the Hamiltonians of
the individual tops, and $H_{12}(t)$ is the coupling between the tops
using spin-spin interaction term with a coupling strength of
$\epsilon/j$. All these angular momentum operators obey standard
commutation relations. For the rest of the paper we will only
concentrate to the case $p_1 = p_2 = \pi/2$. This special choice of
the angular frequencies will simplify both the quantum and classical
maps. Since $J_{i}^{2}$ and $J_{z_i}$'s are four mutually commuting
operators, the simultaneous eigenvectors of these operators we take as
our basis. In general, this basis is denoted by $|j_1, m_1; j_2, m_2
\rangle = |j_1, m_1 \rangle \otimes |j_2, m_2\rangle$, where
$J_{i}^{2} |j_i, m_i\rangle = j_i(j_i+1) |j_i, m_i\rangle$ and $J_{z_i}
|j_i, m_i \rangle = m_i |j_i, m_i\rangle$. The individual top angular
momentums, $j_1$ and $j_2$, could in general be different.

The time evolution operator, defined in between two consecutive kicks,
corresponding to this coupled Hamiltonian is given by,
\begin{equation}
U_T = U_{12}^{\epsilon} (U_1 \otimes U_2) = U_{12}^{\epsilon} \left[ (U_{1}^{k}
U_{1}^{f} \otimes (U_{2}^{k} U_{2}^{f}) \right],
\label{U_T}
\end{equation}
\noindent where the different terms are given by,
\begin{eqnarray}
U_{i}^{f} \equiv \exp\left(-i \frac{\pi}{2} J_{y_i}\right)&;&~~U_{i}^{k} 
\equiv \exp\left(-i\frac{k}{2j} J_{z_i}^{2}\right),\nonumber\\    
U_{12}^{\epsilon} &\equiv& \exp\left(-i\frac{\epsilon}{j} J_{z_1} J_{z_2} 
\right),
\label{Upart}
\end{eqnarray}
\noindent and as usual $i = 1, 2$.

\subsection{\label{CT}Classical Top}

The corresponding classical map of the coupled kicked tops discussed
above can be obtained from the quantum description with the Heisenberg
picture in which the angular momentum operators evolve as:
\begin{equation}
{\bf J}(n+1) = U_{T}^{\dagger} {\bf J}(n) U_T.
\end{equation}
\noindent Now we have to determine the explicit form of this angular momentum
evolution equation for each component of the angular momentum. Here we present
the time-evolution of $J_{x_1}$ (see Appendix \ref{a}):
\begin{eqnarray}
J_{x_1}^{\prime} \equiv U_{T}^{\dagger} J_{x_1} U_{T} &=& \frac{1}{2}
(J_{z_1} + i J_{y_1}) \exp\left[i \frac{k}{j} \left(-J_{x_1} + \frac{1}{2}
\right)\right] \otimes \exp\left(-i \frac{\epsilon}{j} J_{x_2}\right) 
\nonumber\\
&+& \frac{1}{2} \exp\left[ -i \frac{k}{j} \left(-J_{x_1} + \frac{1}{2}\right)
\right] (J_{z_1} - i J_{y_1}) \otimes \exp\left(i \frac{\epsilon}{j} J_{x_2}
\right).
\label{Jx1}
\end{eqnarray}

\noindent The above expression differs from the coupled tops map presented in 
a previous publication \cite{sarkar1}. Firstly, $J_{x_1}^{\prime}$ is
now really a Hermitian operator. Secondly, the terms which arise in
the above expression due to the interaction, contain $J_{x_2}$
operator instead of $J_{y_2}$. We proceed by rescaling the angular
momentum operator as $( X_i, Y_i, Z_i ) \equiv ( J_{x_i}, J_{y_i},
J_{z_i} )/j$, for $i = 1, 2$.  Components of this rescaled angular
momentum vector satisfy the commutation relations, $[X_i,Y_i] =
iZ_i/j, [Y_i,Z_i] = iX_i/j$ and $[Z_i,X_i] = iY_i/j$.  Therefore, in
$j \rightarrow \infty$ limit, components of this rescaled angular
momentum vector will commute and become classical $c$-number
variables.  Consequently, in this large-$j$ limit, we obtain the
classical map corresponding to coupled kicked tops as,
\begin{subequations}
\begin{eqnarray}
X_{1}^{\prime} &=& Z_1 \cos \Delta_{12} + Y_1 \sin \Delta_{12}\\
Y_{1}^{\prime} &=& - Z_1 \sin \Delta_{12} + Y_1 \cos \Delta_{12}\\
Z_{1}^{\prime} &=& - X_1\\
X_{2}^{\prime} &=& Z_2 \cos \Delta_{21} + Y_2 \sin \Delta_{21}\\
Y_{2}^{\prime} &=& - Z_2 \sin \Delta_{21} + Y_2 \cos \Delta_{21}\\
Z_{2}^{\prime} &=& - X_2
\end{eqnarray}
\end{subequations}
\noindent where
\begin{equation}
\Delta_{12} \equiv k X_1 + \epsilon X_2 ; ~~\mbox{and}~~ \Delta_{21} 
\equiv k X_2 + \epsilon X_1.
\end{equation}
\noindent The difference between the map presented above and which was 
derived in \cite{sarkar1} lies in the form of the angles $\Delta_{12}$
and $\Delta_{21}$. However, these differences are very important. The
above map is canonical. It satisfies all Poisson bracket relations
like $\{ X_{i}^{\prime}, Y_{i}^{\prime}\} = Z_{i}^{\prime}, \{
Y_{i}^{\prime}, Z_{i}^{\prime} \} = X_{i}^{\prime}$ and $\{
Z_{i}^{\prime}, X_{i}^{\prime} \} = Y_{i}^{\prime}$, where $i = 1, 2$
; and Poisson brackets of any two dynamical variables corresponding to
different tops are equal to zero. In contrast the classical map
presented in \cite{sarkar1}, satisfies the first three Poisson bracket
relations, but the Poisson brackets of any two dynamical variables
corresponding to different tops are nonzero and they are proportional
to the coupling strength $\epsilon$, implying that the map is
canonical only in the uncoupled limit. Moreover, this earlier
publication relates the entanglement rate to the sum of the positive
Lyapunov exponents, which were actually determined using the incorrect
classical map. However, they considered very weak coupling $(\epsilon
= 10^{-3})$ among the tops and therefore the error in the calculation
of the Lyapunov exponents was very small, these being practically
those of the uncoupled tops. Hence we believe that the main
conclusions presented in that paper are still valid.
   
\begin{figure}
\centerline{\includegraphics[width=3.5in]{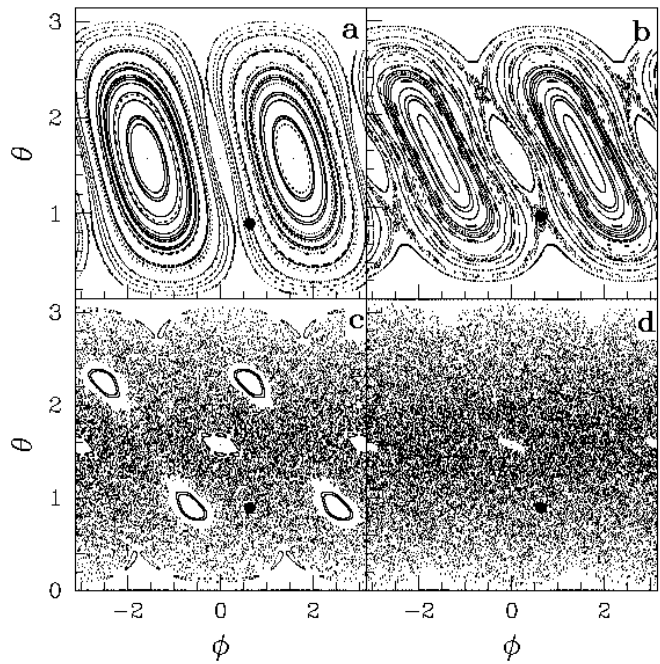}}
\caption{\label{fig1}Phase space pictures of the single top, corresponding to
different parameter values, are presented. (a) $k=1.0$. The phase space is
mostly regular. (b) $k=2.0$. The phase space is still very much regular, but
now a thin stochastic layer is visible at the separatrix. (c) $k=3.0$. The
phase space is truly mixed type. Regular elliptic islands are visible inside
the chaotic region. (d) $k=6.0$. The phase space is mostly covered by the
chaotic region with few tiny elliptic islands. The solid circle is the
point at which we will construct the initial wavepacket.}
\end{figure}

In the limit $\epsilon \rightarrow 0$, we will arrive at the map corresponding
to the single kicked top, whose Hamiltonian is given in Eq. (\ref{single}), and
that map is given by,
\begin{subequations}
\begin{eqnarray}
X^{\prime} &=& Z \cos kX + Y \sin kX \\
Y^{\prime} &=& -Z \sin kX + Y \cos kX \\
Z^{\prime} &=& -X.
\end{eqnarray} 
\end{subequations}
\noindent The classical dynamics of the single top have been studied 
extensively in Ref. \cite{haake1,dariano} and is a well studied model
of quantum chaos. From the above expressions, it is clear that the
variables $(X, Y, Z)$ lie on the sphere of radius unity, i.e., $X^2 +
Y^2 + Z^2 = 1$. This constraint on the dynamical variables restricted
the classical motion to the two-dimensional surface of a unit
sphere. Following the usual procedure, we can parameterize the
dynamical variables in terms of the polar angle $\theta$ and the
azimuthal angle $\phi$ as $X = \sin \theta \cos \phi$, $Y = \sin
\theta \sin \phi$ and $Z = \cos \theta$. In terms of this new
$(\theta, \phi)$ variables, the above map looks very complicated, and
therefore we do not display that map. Moreover, during our numerical
iterations we use the above three-dimensional form of the
map, and after every iteration we get back the corresponding
$(\theta, \phi)$ from the relations $\cos \theta = Z$ and $\phi =
\tan^{-1} (Y/X)$, where $\cos \theta$ and $\phi$ are the canonical
coordinates on the sphere. In Fig. \ref{fig1}, we have presented the
phase-space diagrams of the single top for different values of the
parameter $k$. For $k = 1.0$ and $k = 2.0$, the phase-space is mostly
occupied by regular orbits. As we further increase the value of
$k$, we can see the well known KAM scenario. Finally at $k = 6.0$, the
phase-space is mostly covered by the chaotic sea, with very tiny
islands. The dark circle, marking the point $(\theta, \phi) = (0.89,
0.63)$ in all the phase-space diagrams, is representing the point at
which we will construct our initial wavepacket. The quantities presented
in all the figures are dimensionless. 
\par

\subsection{\label{initial}Initial wavepacket}

We use a generalized $SU(2)$ coherent state or the directed angular momentum 
state \cite{haake1} as our initial state for the individual tops and this  
state is given in $|j, m_i \rangle$ basis as :
\begin{equation}
\langle j, m_i | \theta_{0}, \phi_0 \rangle = ( 1 + |\gamma|^2 )^{-j} 
\gamma^{j - m_i} \sqrt{\left(\begin{array}{c} 2 j\\ j + m_i \end{array} 
\right)},
\label{costate}
\end{equation}
\noindent where $\gamma \equiv \exp(i \phi_0) \tan (\theta_0/2)$. For the
coupled top, we take the initial state as the tensor product of the 
directed angular momentum state corresponding to individual tops. Now on,
we will write $|j, m_i\rangle$ as $|m_i\rangle$ for notational simplification. 
Explicitly in $|m_i\rangle$ basis this initial product state can be written as 
\cite{sarkar1}:
\begin{eqnarray}
|\psi(0) \rangle &=& \sum_{m_1,m_2 = -j}^{+j} \langle m_1, m_2 | \psi(0)\rangle
|m_1, m_2\rangle \nonumber\\
&=& \sum_{m_1,m_2 = -j}^{+j} \langle m_1 | \theta_{0}^{1}, \phi_{0}^{1} \rangle
\langle m_2 |\theta_{0}^{2}, \phi_{0}^{2} \rangle | m_1, m_2 \rangle,
\end{eqnarray}
\noindent where $\langle m_i | \theta_{0}^{i}, \phi_{0}^{i}\rangle,~i = 1, 2$, 
can be obtained from Eq. (\ref{costate}).  

Now we have the evolution $|\psi(n)\rangle = U_{T} |\psi(n-1)\rangle = 
U_{T}^{2} |\psi(n-2)\rangle = .... = U_{T}^{n} |\psi(0)\rangle$. Even though,
the numerical iteration scheme for the above evolutions have already been 
presented in Ref. \cite{sarkar1}, here we again present that for the sake
of completeness. From \cite{sarkar1}, we have 
\begin{equation}
\langle s_1, s_2 | \psi(n) \rangle = \exp \left(-i\frac{\epsilon}{j} s_1 s_2
\right) \sum_{m_1,m_2 = -j}^{+j} \langle s_1 | U_1 |m_1 \rangle \langle s_2 |
U_2 | m_2 \rangle \langle m_1, m_2 | \psi(n-1)\rangle
\end{equation}
\noindent where
\begin{equation}
\langle s_1 | U_1 | m_1 \rangle = \exp \left( -i \frac{k}{2j} s_{1}^{2} \right)
d_{s_1 m_1}^{(j)} \left(\frac{\pi}{2}\right).
\end{equation}
\noindent $d_{s_1 m_1}^{(j)} \left(\frac{\pi}{2}\right)$ is the Wigner rotation
matrix \cite{sakurai} :
\begin{equation}
d_{s_1 m_1}^{(j)} \left(\frac{\pi}{2}\right) =
\frac{(-1)^{s_1-m_1}}{2^j} \left(\begin{array}{c} 2j\\j-s_1 \end{array}
\right)^{1/2} \left(\begin{array}{c} 2j\\j+m_1 \end{array} \right)^{-1/2}
\sum_{k} (-1)^k \left( \begin{array}{c} j-s_1\\k \end{array} \right) \left(
\begin{array}{c} j+s_1\\k+s_1-m_1 \end{array}\right).
\label{wigner}
\end{equation}
\noindent The main problem in calculating the Wigner rotation matrix lies
in the calculation of the above sum. Defining that sum as $V_{m_1}$, and
starting from $V_{-j} = 1$ and $V_{- j + 1} = 2\,s_1$, we can get the other
$V_{m_1}$ recursively by using the following relation \cite{peres}
\[ ( j - m_1 + 1 ) V_{m_1 - 1} - 2 s_1 V_{m_1} + ( j + m_1 + 1 ) V_{m_1 + 1}
= 0. \]
Besides Wigner rotation matrix can be expressed in terms of Jacobi 
polynomials and of different Hypergeometric functions \cite{varshal}. 
However, we have followed the above recursive scheme.

\subsection{\label{SvSr}Measures of entanglement}

All the previous studies on the connection between entanglement and chaos,
were based on pure states of bipartite system, where the von Neumann entropy
$S_V$ and the Linear entropy $S_R$ of the reduced density matrices (RDMs) were
natural measures of entanglement. The definition of these entropies are:
\begin{eqnarray}
S_V (n) &=& - \mbox{Tr}_1 [ \rho_1(n) \ln \rho_1 (n) ] = - \mbox{Tr}_2
[ \rho_2(n) \ln \rho_2 (n) ]\\
\mbox{and}~~S_R(n) &=& 1 - \mbox{Tr}_1 \left[ \rho_{1}^{2}(n)\right] = 1 -
\mbox{Tr}_2 \left[ \rho_{2}^{2}(n) \right]
\end{eqnarray}
\noindent where $\rho_1$ and $\rho_2$ are the RDMs corresponding to the first
and the second top respectively. In the eigenbasis of the RDM :
\begin{eqnarray}
S_V (n) &=& - \sum_{i} \lambda_i \ln \lambda_i \\
S_R (n) &=& 1 - \sum_{i} \lambda_{i}^{2},
\end{eqnarray}
\noindent where $\lambda_i$'s are the eigenvalues of the RDMs.

\subsection{\label{redhus}Reduced Husimi function}

Since the phase space of the coupled tops is four dimensional $(S^2 \times 
S^2$), it is not possible to visualize the wavepacket dynamics on such a  
phase space. Therefore, we use 
an approximate numerical way to visualize the behavior of the time evolving
state $|\psi(n)\rangle$ in any one of its subspaces. We call this method 
{\it reduced Husimi function} and it is defined in the following way. Consider 
a state $|\psi\rangle$ in the angular momentum basis $|m_1, m_2\rangle$, i.e.,
\begin{equation}
|\psi\rangle = \sum_{m_1, m_2} a_{m_1m_2} |m_1, m_2\rangle.
\end{equation}
\noindent The Husimi function of $|\psi\rangle$ is $|\langle z_1; z_2 | \psi
\rangle |^2$, where 
\begin{equation}
\langle z_1 ; z_2 | \psi\rangle = \sum_{m_1, m_2}
a_{m_1m_2} \langle z_1 | m_1\rangle \langle z_2 | m_2 \rangle,
\end{equation}
\noindent and $|z_i\rangle \equiv |\theta_i, \phi_i \rangle$ are the directed
angular momentum states (atomic coherent states). We define reduced Husimi 
function corresponding to first subspace,
\begin{equation}
\rho_{1H} (z_1) = \int_{z_2} |\langle z_1 ; z_2 | \psi\rangle|^2 d\mu(z_2),
\label{rho1H}
\end{equation}
\noindent where $d\mu(z_2)$ is the Haar measure :
\begin{equation}
d\mu(z_2) = \frac{2 j + 1}{4 \pi} \sin \theta_2 d\theta_2 d\phi_2.
\end{equation}
Since the phase space of a kicked top is the surface of a sphere of unit
radius, the total phase space area is $4 \pi$. Therefore for the 
kicked top whose Hilbert space dimension is $N = 2 j + 1$, volume of the
Planck cell is $4 \pi/ (2 j + 1)$. Hence the above mentioned Haar measure
$d \mu (z)$ is equal to the number of Planck cells present in the 
infinitesimal area $d z = \sin \theta d \theta d \phi$. The integration
of $d \mu (z)$ over whole phase space will give total number of Planck cells
$N = 2 j + 1$ present in the whole phase space. One can also write the above 
expression, Eq. (\ref{rho1H}), as,


\begin{equation}
\rho_{1H} (\theta_1, \phi_1) = \left\langle \theta_1, \phi_1 \left|
\frac{2j+1}{4\pi} \left[ 
\int_{\theta_2} \int_{\phi_2} \langle \theta_2, \phi_2 | \psi \rangle \langle 
\psi | \theta_2, \phi_2 \rangle \sin\theta_2 d\theta_2 d\phi_2 \right] \right| 
\theta_1, \phi_1 \right\rangle.
\end{equation}
\noindent The above integral is just the partial trace of the density matrix 
$|\psi\rangle \langle \psi|$ over the second subspace, and hence it gives the 
reduced density matrix (RDM) corresponding to the first subspace. Therefore,
\begin{equation}
\rho_{1H}(\theta_1, \phi_1) = \langle \theta_1, \phi_1 | \rho_1 | \theta_1, 
\phi_1 \rangle,
\end{equation}
\noindent where $\rho_1$ is the RDM of the first subspace. Therefore, the
reduced Husimi function is just the Husimi function of the RDM. We can write
$\rho_1 = \sum_{i=1}^{N} \lambda_i |e_i\rangle \langle e_i |$, where 
$\lambda_i$'s are the eigenvalues of $\rho_1$ and $|e_i\rangle$'s are the 
corresponding eigenstates. These $|e_i\rangle$'s are also called Schmidt 
vectors. Therefore,
\begin{equation}
\rho_{1H}(\theta_1, \phi_1) = \sum_{i=1}^{N} \lambda_i \left|\langle\theta_1, 
\phi_1 | e_i \rangle \right|^2.
\end{equation}
\noindent Thus the reduced Husimi function can also be expressed as the 
weighted sum of the Husimi
functions of the Schmidt vectors, where the weight factors are the eigenvalues
of the RDM. Identically, we can define reduced Husimi function for the second
subspace, and is given by,
\begin{equation}
\rho_{2H}(\theta_2, \phi_2) = \sum_{i=1}^{N} \lambda_i \left|\langle\theta_2, 
\phi_2 | d_i \rangle \right|^2,
\end{equation}
\noindent where $|d_i\rangle$'s are the Schmidt vectors of the second 
subspace.

\section{\label{W2}Second moment of Husimi function : a measure of complexity
of a quantum state}

Reduced Husimi function technique is useful for the visualization of the 
behavior of the time evolving state on the phase space. Moreover, we want a 
phase space measure of the complexity of any state to relate it with the 
entanglement. There already exists a good measure of that complexity
based on the Husimi distribution function, $\rho_H = \langle z | \rho | z 
\rangle$, called `classical entropy' or Wehrl entropy \cite{wehrl} and that 
is given by
\begin{equation}
S(\rho_H) = \int d\mu(z) \rho_H \ln \rho_H
\end{equation}
\noindent However, it is difficult to determine the above quantity due to the
presence of the logarithmic function. Therefore, following a recent proposal 
\cite{sugita}, we consider inverse of the `second moment of the 
Husimi function' $W_2(\rho_H)$ as a measure complexity of quantum states. 
This measure is defined as,
\begin{eqnarray}
W_2(\rho_H) &=& \frac{1}{M_2(\rho_H)}\\
\mbox{where}~~M_2(\rho_H) &=& \int d\mu(z) \rho_{H}^{2}.
\end{eqnarray}
\noindent The quantity $W_2$ represents the effective phase space occupied by 
the Husimi function of the state $\rho$ and its unit is the Planck's
cell volume. We note that a similar kind of quantity, based on the
Wigner function, has been introduced and studied as a measure of the
complexity of quantum states in phase space \cite{heller} many years ago.

We can now define a quantity $\Delta N_{\mbox{eff}} \equiv W_2(\rho_H)/N$
as the fraction of the total number of Planck cells $( N = 2 j + 1 )$
occupied by the state $\rho$. Since the total number of Planck cells is 
equal to the Hilbert space dimension, we can define $\Delta N_{\mbox{eff}}$
also as the rough measure of the fraction of the Hilbert space occupied
by the above state. The above definitions of $\Delta N_{\mbox{eff}}$ are 
valid for the single top. For the coupled tops, phase space is 
$4$-dimensional. Here, we can define $\Delta N_{\mbox{eff}}$ for 
any one of its subspaces. However, the only difference between these two 
cases is that $\rho$ is a pure state for the single top whereas for the
coupled tops, $\rho$ is a mixed state. Here we have studied the time evolution 
of $\Delta N_{\mbox{eff}}$ for the single top and also for the coupled tops.

\subsection{\label{singletop}Single top}

In the single top case, we have again considered $SU(2)$ coherent state
$|\psi(0)\rangle = |\theta_0,\phi_0\rangle$, which we have already defined
in Eq.(\ref{costate}), as the initial state. We have constructed this state at
the point $(\theta_0,\phi_0) = (0.89, 0.63)$, and evolved it with repeated
applications of the single top evolution operator $U$. The time evolution
operator $U$, defined between two consecutive kicks, is given as
\begin{equation}
U = \exp\left(-i\frac{\pi}{2} J_y\right)\exp\left(-i\frac{k}{2j} J_{z}^{2}
\right).
\end{equation}
\noindent For the single top case, $\Delta N_{\mbox{eff}}$ at time $n$ is
\begin{eqnarray}
\Delta N_{\mbox{eff}} &=& \frac{1}{(2j+1) M_2[|\psi(n)\rangle]}\nonumber\\
\mbox{where}~~~M_2[|\psi(n)\rangle] &=& \int d\mu(z) |\langle z | \psi(n)
\rangle|^4
\end{eqnarray}
\par
\begin{figure}
\centerline{\includegraphics[width=3.5in]{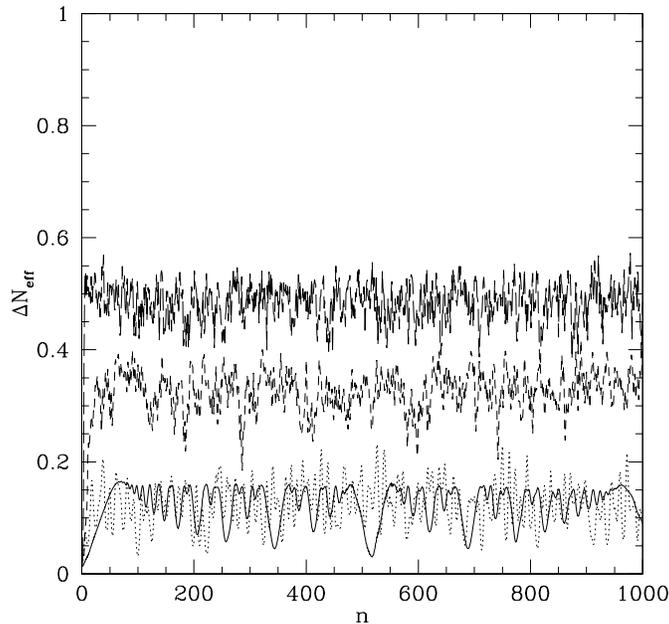}}
\caption{\label{fig2} Evolution of $\Delta N_{\mbox{eff}}$ is presented for
the single top. For the nonchaotic cases ( $k = 1.0$ and $k = 2.0$ ),
denoted respectively by solid and dotted line, maximum value of
$\Delta N_{\mbox{eff}}$ is very less. That means, the time evolving
state has very little access over the phase space. However, for
chaotic cases ( $k = 3.0$ and $k = 6.0$ ), maximum value of $\Delta
N_{\mbox{eff}}$ is also not large. For the strongly chaotic case ( $k =
6.0$ ), the average value of the maxima is about $0.5$.}
\end{figure}
\par
\noindent and $|\psi(n)\rangle = U^n |\psi(0)\rangle$. In Fig.\ref{fig2},
we have shown time evolution of $\Delta N_{\mbox{eff}}$ for different
$k$-values. For $k=1.0$, the initial state is inside the elliptic
region, and therefore, time evolution of this state is governed by the
elliptic orbits on which it is initially placed. Since the evolution
of this state is in some sense trapped by the elliptic orbits, it has
little or no access to many parts of the phase space.  Consequently,
the maximum value of $\Delta N_{\mbox{eff}}$ is very small. After
reaching its maxima, there are many oscillations in the time evolution
of $\Delta N_{\mbox{eff}}$ due to partial and full revival of the time
evolving state $|\psi(n)\rangle$. This particular issue of quantum
revival of the time evolving state in such mixed systems warrant a
separate study. Now at $k=2.0$, the initial state is inside a
stochastic layer present at the separatrix and consequently its
dynamics is restricted by and large to be inside that stochastic
layer. Naturally, for this case, the maxima of $\Delta N_{\mbox{eff}}$
is again small. For $k=3.0$, the phase space is of a truly mixed type, with
a significant measure of chaotic orbits. In this case, the initial state is
inside the chaotic region.  Therefore, time evolution of this state is
governed by the chaotic dynamics and this state has access over
chaotic region of the phase space. Since the size of the chaotic
region is large, hence the maxima of $\Delta N_{\mbox{eff}}$ is larger
$(\sim 0.35)$. When $k=6.0$, the phase space is mostly covered by the
chaotic region, with few visible tiny regular islands. The time
evolving state has now almost full access over the phase
space. However, we observed that $\Delta N_{\mbox{eff}}$ reaches
maximum around $0.5$ and then fluctuates around that value. That
means, for this strong chaotic case, the time evolving {\it pure} state has
access over only half of the phase space. This typical behavior of
$\Delta N_{\mbox{eff}}$ for strongly chaotic case can be explained by
RMT in the following way.

In the angular momentum basis $\{|m\rangle\}$,
\begin{equation}
M_2[|\psi(n)\rangle] = \sum_{i,k} \sum_{l,m} \langle i | \psi(n) \rangle
\langle \psi(n) | k \rangle \langle l | \psi(n) \rangle \langle \psi(n)
| m \rangle \int d\mu(z) \langle z | i \rangle \langle k | z \rangle \langle 
z | l \rangle \langle m | z \rangle.
\label{integral}
\end{equation}
\noindent After performing the above integral, see Appendix \ref{c},
\begin{eqnarray}
M_2\left[|\psi(n)\rangle\right] &=& \sum_{i,k} \sum_{l,m}
\langle i | \psi(n) \rangle \langle \psi(n) | k \rangle
\langle l | \psi(n) \rangle \langle \psi(n) | m \rangle F(2 j; i, k, l, m)
\delta_{i+l,k+m} \label{sol1}\\
\mbox{where}~~~ F(2 j; i, k, l, m) &=&  \frac{2 j + 1}{(4 j + 1)!}
\sqrt{\left(\begin{array}{c}2 j\\j - i\end{array}\right)
\left(\begin{array}{c}2 j\\j - k\end{array}\right)
\left(\begin{array}{c}2 j\\j - l\end{array}\right)
\left(\begin{array}{c}2 j\\j - m\end{array}\right)} (2j-i-l)! (2j+i+l)!
\label{F_func}
\end{eqnarray}

Let us assume, in the angular momentum basis,
\begin{equation}
|\psi(n)\rangle = \sum_m c_m |m\rangle.
\end{equation}
\par
\begin{figure}
\centerline{\includegraphics[width=3.5in]{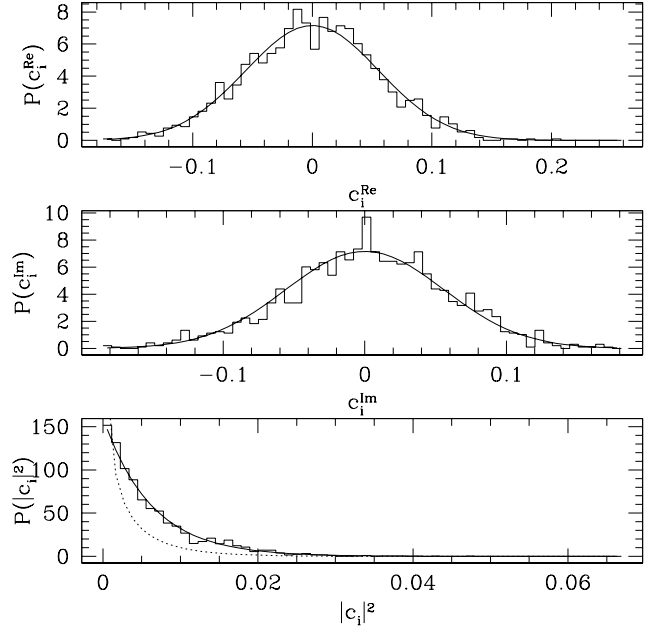}}
\caption{\label{fig3}Distribution of the components of the time evolving state,
evolving under strongly chaotic single top dynamics, is presented. Top and
middle windows are showing that the real and the imaginary part of the
components of the time evolving state are Gaussian distributed random numbers
with {\it zero} mean and the variance is $1/\sqrt{N}$, where $N = 2 j + 1$ is
the Hilbert space dimension of the top. In this case $j = 80$. Bottom window is
showing that the distribution of the square of the absolute values of the
components of the time evolving state are exponentially distributed. This is a
typical property of the components of a GUE distributed vector. Dotted line
representing the GOE (Porter-Thomas) distribution.}
\end{figure}
\par
\noindent In Fig.\ref{fig3}, we have presented the distribution of the real
and the imaginary part of the coefficients $c_m$. They are indeed Gaussian
distributed random numbers. Moreover, in that figure, we have also presented
the distribution of $|c_m|^2$. This figure shows that $|c_m|^2$ are
exponentially distributed, which is a typical property of the elements of a
Gaussian unitary ensemble (GUE) distributed random vector. Therefore, we can
assume that the distribution of $\{c_m\}$ are GUE type. For GUE case, RMT
average of a quantity identical to $M_2[|\psi(n)]$ has been calculated in a 
recent publication \cite{manderfeld}, and according to that, 
\begin{equation}
\langle M_2[|\psi(n)]\rangle = \frac{2}{N+1}, ~~\mbox{where}~~N = 2j+1,
\end{equation}
\noindent where the angular bracket $\langle~\rangle$  represents RMT
averaged value. Using the above expression, we have
\begin{equation}
\langle \Delta N_{\mbox{eff}} \rangle = \frac{N+1}{2N} = \frac{1}{2}
\left( 1 + \frac{1}{N} \right)
\end{equation}
\noindent and for large $N$ limit,
\begin{equation}
\langle \Delta N_{\mbox{eff}} \rangle \simeq 0.5.
\end{equation}
\noindent This is the saturation value of $\Delta N_{\mbox{eff}}$, which was
observed in strongly chaotic case $k=6.0$.

\subsection{\label{coupDelN}Coupled tops}

In the last section, we have presented reduced Husimi function technique to
visualize the behaviors of the time evolving state of the coupled tops on any
one of its subspaces. However, to measure the complexity of this state in any
one of its subspaces, we have to define $\Delta N_{\mbox{eff}}$ in a subspace.
We have defined $\Delta N_{\mbox{eff}}$ for a given subspace as
\begin{eqnarray}
\Delta N_{\mbox{eff}} &=& \frac{1}{(2j+1) M_2(\rho_{iH})}\\
\mbox{where}~~~M_2(\rho_{iH}) &=& \int d \mu(z_i)
\langle z_i | \rho_i | z_i \rangle,
\end{eqnarray}
\par
\begin{figure}
\centerline{\includegraphics[height=3.5in]{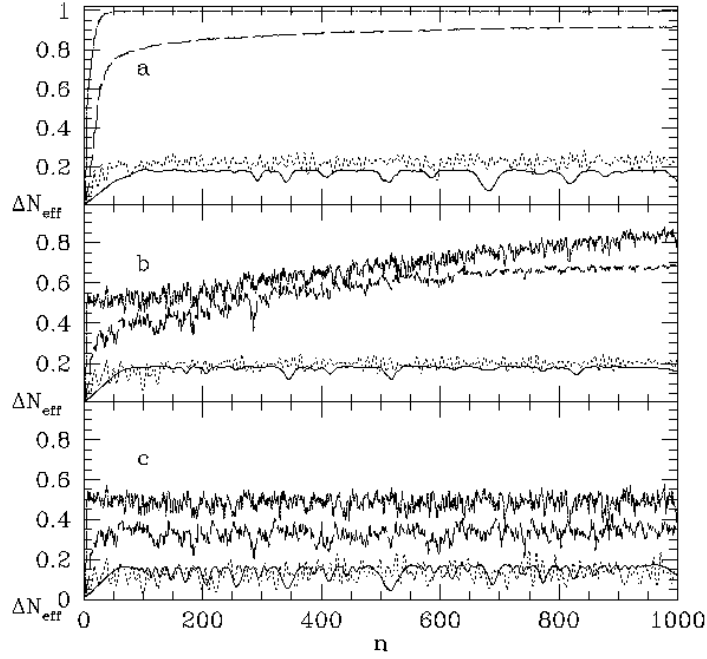}}
\caption{\label{fig4}Evolution of $\Delta N_{\mbox{eff}}$ corresponding to
coupled kicked tops is presented. Solid lines and dotted lines are representing
the results corresponding to nonchaotic cases ( $k = 1.0$ and $k = 2.0$
respectively ). Dashed lines are representing the mixed case ( $k = 3.0$ ) and
dash-dot lines are showing the results corresponding to strongly chaotic case
( $k = 6.0$ ). The top window representing the results for the stronger
coupling strength ( $\epsilon = 10^{-2}$ ), middle window is showing the
results for the intermediate coupling strength ( $ \epsilon = 10^{-3}$ ) and
the bottom window is for the weak coupling case ( $\epsilon = 10^{-4}$ ).}
\end{figure}
\par
\noindent and $i = 1, 2$ representing different subspaces. In Fig.\ref{fig4},
we have presented the time evolution of the above mentioned $\Delta
N_{\mbox{eff}}$ for different dynamics (different $k$ values) and for
different coupling strengths $\epsilon$. When coupling strength is
very weak $(\epsilon = 10^{-4})$, time evolution of $\Delta
N_{\mbox{eff}}$ for different dynamics are practically identical to
that which we have observed in the case of single tops. Therefore, for this
coupling strength, effect of the dynamics of one top on the other top
is very small and two tops are very close to two uncoupled
systems. For other coupling strengths, the maxima of $\Delta
N_{\mbox{eff}}$ has not changed much for the nonchaotic cases
$(k=1.0,$ and $k=2.0)$. When $\epsilon = 10^{-3}$, for the chaotic
cases $(k=3.0,$ and $k=6.0)$, $\Delta N_{\mbox{eff}}$ first reaches
the saturation value which is observed in the case of single tops and
then it increases approximately linearly with time. However, for the
stronger coupling $(\epsilon = 10^{-2})$, it is not possible to divide
the time evolution of $\Delta N_{\mbox{eff}}$, for the chaotic cases,
into two distinct time regimes. In these cases, $\Delta
N_{\mbox{eff}}$ saturates at much higher values than the maxima of
$\Delta N_{\mbox{eff}}$ observed in single top. For the strongly chaotic
case $k=6.0$, $\Delta N_{\mbox{eff}}$ saturates at a value that is
very slightly less than unity. This saturation of $\Delta N_{\mbox{eff}}$ can
be also explained by RMT, which we now proceed to do.

In the angular momentum basis, second moment of the Husimi function of the
reduced state, say for the first subsystem, at time $n$ is,
\begin{equation}
M_2(\rho_{1H}) = \sum_{i,k} \sum_{l,m} (\rho_1)_{ik}^{} (\rho_1)_{lm}^{} \int
d\mu(z_1) \langle z_1 | i \rangle \langle k | z_1\rangle \langle z_1 | l
\rangle \langle m | z_1 \rangle.
\label{M2rho1H}
\end{equation}
\noindent After performing the above integral, we have,
\begin{equation}
M_2(\rho_{1H}) = \sum_{i,k} \sum_{l,m} (\rho_1)_{ik} (\rho_1)_{lm}
F(2 j ; i, k, l, m) \delta_{i+l,k+m}
\label{sol2}
\end{equation}
\noindent where $F(2 j ; i, k, l, m)$ has already been given in Eq.
(\ref{F_func}). If we write down above expression in the eigenbasis of the RDM
$\rho_1$, then we have,
\begin{eqnarray}
M_2(\rho_{1H}) &=& \sum_{\alpha,\beta = 1}^{N} \lambda_{\alpha} \lambda_{\beta}
\sum_{i,k} \sum_{l,m} \langle i | \phi_{\alpha} \rangle \langle \phi_{\alpha}
| k \rangle \langle l | \phi_{\beta} \rangle \langle \phi_{\beta} | m \rangle
F(2 j ; i, k, l, m) \delta_{i+l,k+m}\nonumber\\
&=& \sum_{\alpha} \lambda_{\alpha}^{2} \sum_{i,k} \sum_{l,m} \langle i
| \phi_{\alpha} \rangle \langle \phi_{\alpha} | k \rangle \langle l | 
\phi_{\alpha} \rangle \langle \phi_{\alpha} | m \rangle F(2 j ; i, k, l, m) 
\delta_{i+l,k+m} \nonumber\\ 
&+& \sum_{\begin{array}{c} \alpha, \beta \\ \alpha \neq \beta \end{array}} 
\lambda_{\alpha} \lambda_{\beta} \sum_{i,k} \sum_{l,m} \langle i | 
\phi_{\alpha} \rangle \langle \phi_{\alpha} | k \rangle \langle l | 
\phi_{\beta} \rangle \langle \phi_{\beta} | m \rangle  
F(2 j ; i, k, l, m) \delta_{i+l,k+m}\nonumber\\
&\equiv& \sum_{\alpha} \lambda_{\alpha}^2 Q_{\alpha\alpha}^2 +
\sum_{\begin{array}{c} \alpha, \beta \\ \alpha \neq \beta \end{array}}
\lambda_{\alpha} \lambda_{\beta} Q_{\alpha\beta}^2
\end{eqnarray}
\begin{eqnarray}
\mbox{where}~~Q_{\alpha\alpha}^2 &=& \sum_{i,k} \sum_{l,m} \langle i |
\phi_{\alpha} \rangle \langle \phi_{\alpha} | k \rangle \langle l |
\phi_{\alpha} \rangle \langle \phi_{\alpha} | m \rangle F(2 j; i, k, l, m),\\
\mbox{and}~~Q_{\alpha\beta}^2 &=& \sum_{i,k} \sum_{l,m}\langle i |
\phi_{\alpha} \rangle \langle \phi_{\alpha} | k \rangle \langle l |
\phi_{\beta} \rangle \langle \phi_{\beta} | m \rangle F(2 j; i, k, l, m),
\end{eqnarray}
\noindent where $\{ \lambda_{\alpha}, | \phi_{\alpha} \rangle \}$ are the
eigenvalues and the eigenvectors of the RDM $\rho_1$.
\par
\begin{figure}
\centerline{\includegraphics[height=3.5in]{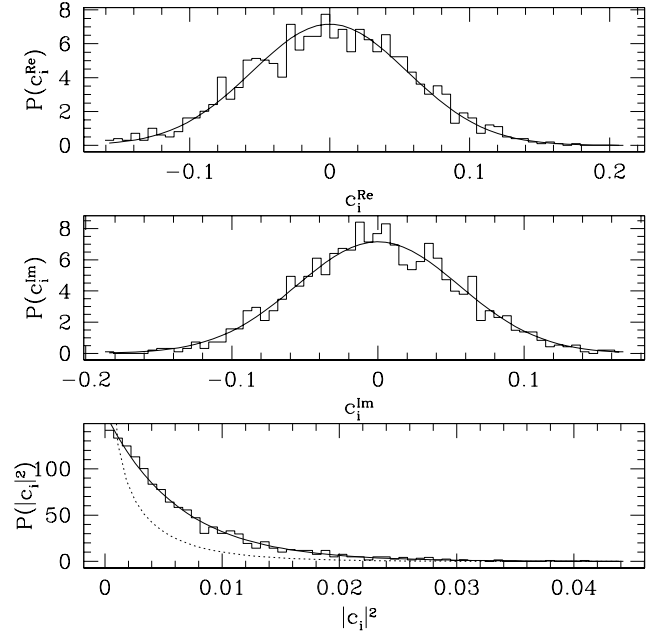}}
\caption{\label{fig5}Distribution of the components of the eigenvectors
of the RDM, corresponding to which entanglement production has reached the
statistical bound. The top and the middle window shows that
the real and the imaginary part of the components of these eigenvectors of
RDM are Gaussian distributed random numbers with {\it zero} mean and the
variance is $1/\sqrt{N}$. Here $N = 2 j + 1 = 161$. The bottom window is
showing that the distribution of the absolute square of the eigenvectors of
the RDM are exponentially distributed. Therefore, the eigenvectors of the RDM
are GUE distributed. Dotted line representing the GOE ( Porter-Thomas )
distribution.}
\end{figure}
\par

In Fig.\ref{fig5}, we have presented the distribution of the real and the
imaginary part of the components of the eigenvectors $\{ |\phi_{\alpha} 
\rangle\}$ of the RDM $\rho_1$. This figure shows that the real and the 
imaginary part of $\{ |\phi_{\alpha} \rangle \}$ are Gaussian distributed 
random numbers. Moreover, Fig.\ref{fig5} also shows that the distribution of 
the absolute square of the components of $\{ |\phi_{\alpha} \rangle \}$ is GUE 
type. Therefore, from the recent calculation \cite{manderfeld}, we can again 
use RMT average values of $Q_{\alpha\alpha}^2$ and $Q_{\alpha\beta}^2$ to get 
RMT average value of $M_2(\rho_{1H})$ as,
\begin{eqnarray}
\langle M_2(\rho_{1H}) \rangle &=& \frac{2}{N+1} \left\langle \sum_{\alpha}
\lambda_{\alpha}^2 \right\rangle + \frac{1}{N+1} \left\langle 
\sum_{\begin{array}{c}
\alpha, \beta\\ \alpha \neq \beta \end{array}} \lambda_{\alpha} \lambda_{\beta}
\right\rangle\nonumber\\
&=& \frac{2}{N+1} \left\langle \sum_{\alpha} \lambda_{\alpha}^2 \right\rangle +
\frac{1}{N+1} \left[ 1 - \left\langle \sum_{\alpha} \lambda_{\alpha}^2 
\right\rangle \right]\nonumber\\
&=& \frac{1}{N+1} \left( 1 + \left\langle \sum_{\alpha} \lambda_{\alpha}^2 
\right\rangle \right).
\end{eqnarray}
\noindent We know from our earlier work \cite{our},
\begin{equation}
\left\langle \sum_{\alpha} \lambda_{\alpha}^2 \right\rangle = 
\frac{2N+1}{N^2+2}.
\end{equation}
\noindent Therefore, we have,
\begin{equation}
\left\langle M_2(\rho_{1H}) \right\rangle = \frac{1}{N+1} \left( 1 + 
\frac{2N+1}{N^2+2} \right).
\end{equation}
\noindent Hence,
\begin{eqnarray}
\left\langle \Delta N_{\mbox{eff}}\right\rangle &=& \frac{1}{N \left\langle 
M_2(\rho_{1H}) \right\rangle} \nonumber\\
&=& \frac{(N+1)(N^2+2)}{N (N^2 + 2N + 3)}.
\end{eqnarray}
\noindent In the large $N$ limit,
\begin{equation}
\left\langle \Delta N_{\mbox{eff}}\right\rangle \,=\, \frac{N+1}{N+2} 
\,+\, {\cal O}\left(\frac{1}{N^2}\right)\, \overset{<}{\sim} 1.0.
\end{equation}
\noindent This is the saturation value of $\Delta N_{\mbox{eff}}$, which we
have observed in the strongly chaotic $(k=6.0)$ and strongly coupled
$(\epsilon = 10^{-2})$ case. We emphasize that this is nearly twice
that of pure states in a single top. Thus roughly speaking the effect
of strongly coupling to another chaotic system doubles the phase space
access of a state.

\section{\label{numerics}Numerical Results}

\subsection{Classical Phase space}

In Fig.\ref{fig1}, we have presented the phase space picture of the
single kicked top for different parameter values. For $k = 1.0$, as
shown in Fig.\ref{fig1}(a), the phase space is mostly covered by
regular orbits, without any visible stochastic region. Our initial
wavepacket, marked by a solid circle at the coordinate $(0.89,0.63)$,
is on the regular elliptic orbits. As we further increase the
parameter, regular region becomes smaller.  Fig.\ref{fig1}(b) is
showing the phase space for $k = 2.0$. Still the phase space is mostly
covered by the regular region, but now we can observe a thin
stochastic layer at the separatrix. In this case, the initial
wavepacket is on the separatrix. For the change in the parameter value
from $k=2.0$ to $k=3.0$, there is significant change in the phase
space. At $k=3.0$, shown in Fig.\ref{fig1}(c), the phase space is of a
truly mixed type. The size of the chaotic region is now very large
with few regular islands. At this parameter value, the initial
wavepacket is inside the chaotic region. Fig.\ref{fig1}(d) is showing
the phase space for $k=6.0$. Now the phase space is mostly covered by
the chaotic region, with very tiny regular islands. Naturally, our
initial wavepacket is in the chaotic region.

\subsection{\label{dynamics}Time evolution of the quantum entanglement}

\begin{figure}
\centerline{\includegraphics[height=4.5in]{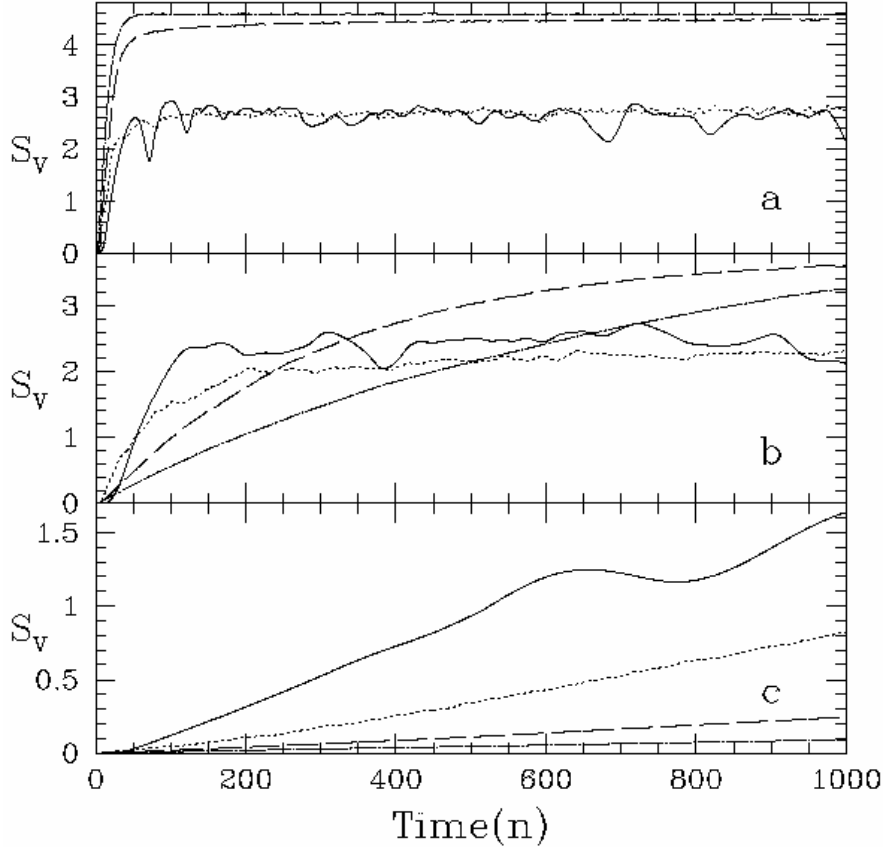}}
\caption{\label{fig6} Time evolution of the von Neumann entropy in coupled
kicked tops is presented for different coupling strengths and for different
underlying classical dynamics. (a) $\epsilon = 10^{-2}$. (b) $\epsilon =
10^{-3}$. (c) $\epsilon = 10^{-4}$. Solid line represents $k=1.0$, dotted line
corresponds to $k=2.0$, dashed line is for $k=3.0$ and dash-dot line
represents $k=6.0$.}
\end{figure}

In Fig.\ref{fig6}, we have presented our results for the entanglement 
production in coupled kicked tops for the spin $j = 80$. As we go from top to 
bottom window, coupling strength is decreasing by a factor of {\it ten}. Top 
window corresponds to $\epsilon = 10^{-2}$, middle one is showing the results 
for $\epsilon = 10^{-3}$ and the bottom window corresponds to the case 
$\epsilon = 10^{-4}$. For each coupling strength, we have studied entanglement 
production for four different single top parameter values, whose corresponding 
classical phase space picture has already been shown in Fig.\ref{fig1}.
\par
\par

\subsubsection{Coupling $\epsilon = 10^{-2}$}

Let us first discuss the case of stronger coupling $\epsilon =
10^{-2}$, whose results are presented in Fig.\ref{fig6}a. It shows
that there exists a saturation of $S_V$ for $k = 1.0$ and $k = 2.0$,
which are much smaller than the saturation value corresponding to
highly chaotic cases such as when $k = 6.0$. The saturation value of
$S_V$ for $k = 6.0$ is the statistical bound $S_V =
\ln(N)-\frac{1}{2} \simeq 4.57$ (where $N = 161$), which can be 
understood from random matrix theory \cite{our}. However for $k =
3.0$, corresponding to a mixed classical phase space, $S_V$ is still
less than the above mentioned saturation value, indicating the
influence of the regular regions.

These two distinct behaviors of the entanglement saturation can be
understood from the underlying classical dynamics. For $k = 1.0$, the
initial unentangled state is the product of the coherent wavepacket
placed inside the elliptic region [see Fig.\ref{fig1}(a)] of each
top. This initially unentangled state will become more and more
entangled under the repeated application of the coupled top unitary
operator $U_T$. Moreover, if one observes the evolution of the reduced
Husimi function corresponding to each top, then it can be seen that
the initially localized wavepacket starts moving along the classical
elliptic orbits on which it was initially placed and simultaneously it
also spreads along those orbits. 

However, one can observe some initial oscillations in the entanglement
production, which is due to the fact that the entanglement production
is mostly determined by the spreading of the wavepacket along
$\theta$-direction. As we know $\cos\theta_i =
\mbox{lim}_{j\rightarrow \infty}(J_{z_i}/j)$, therefore the spreading
of the wavepacket along $\theta$-direction determines how many
eigenstates of $J_{z_i}$, which are also our basis states, are
participating to construct the wavepacket. Larger amount of spreading
of the wavepacket along the $\theta$-direction causes greater number
of basis states to participate in the wavepacket. Moreover, coupling
between two tops is via interaction between $J_{z_1}$ and
$J_{z_2}$. Therefore, this interaction term will couple greater number
of basis states and consequently leads to higher
entanglement. 
\par
\begin{figure}
\centerline{
\begin{tabular}{c}
\includegraphics[height=2.0in]{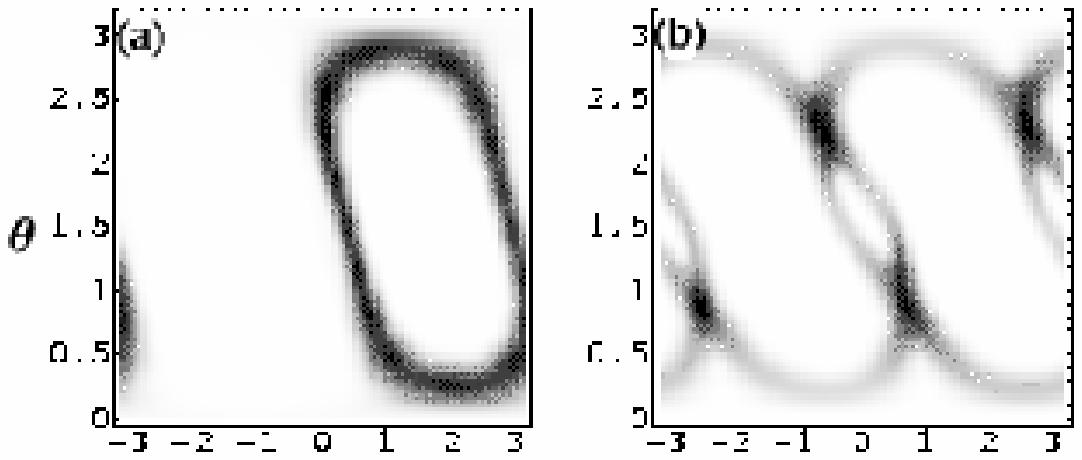}\\
\includegraphics[height=2.0in]{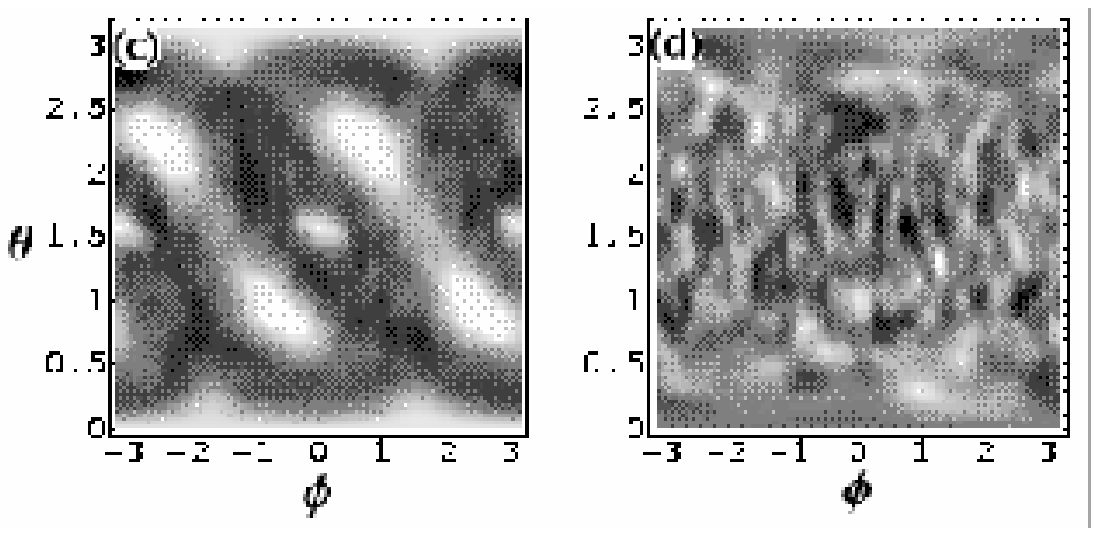}
\end{tabular}
}
\caption{\label{fig7}Reduced Husimi functions of the time evolving state,
evolving under $U_T$,
are presented corresponding to the time at which the entanglement production
is saturated. (a) $k=1.0$. The wavepacket is spread over the elliptic
orbits. (b) $k=2.0$. The wavepacket is spread over the separatrix. It is also
showing strong localization at the unstable period-$4$ orbit. (c) $k=3.0$. The
wavepacket is spread over the whole chaotic region. (d) $k=6.0$. At this
parameter value, the phase space is mostly covered by the chaotic region, see
Fig.\ref{fig1}. Consequently, the wavepacket is spread over almost whole
phase space.}
\end{figure}
\par

Initially, the spreading of the wavepacket sometimes may become
parallel to the $\phi$-direction and therefore its spreading along
$\theta$-direction become less. Consequently, one can observe a dip in
the entanglement production. Finally, the wavepacket spreads all over
the elliptic orbits and the entanglement production reaches its
saturating maxima. In Fig.\ref{fig7}a, we have shown the reduced
Husimi function of the wavepacket corresponding to the maxima
(saturation) of the entanglement production. After reaching its
saturation, there are again many dips in the entanglement
production. These dips are also due to the small spreading of the
wavepacket along $\theta$-direction. However, the localization of the
wavepacket along $\theta$-direction are now happening due to
fractional or full revival of the wavepacket. These revivals are
actually the single top behaviors which persists even under the
interaction with other top. The quantum revivals of the wavepacket are
interesting phenomena of any quantum system and therefore it requires
separate study, especially in this rather more complex setting.

At $k = 2.0$, the center of the initial coherent state was inside the
separatrix. Therefore, in its time evolution, the spreading of the
wavepacket was restricted to be inside the separatrix region. Finally, the
wavepacket spread over the separatrix region, and the entanglement
production arrived at its saturation. The corresponding reduced Husimi
function has been shown in Fig.\ref{fig7}b. Moreover, the reduced
Husimi function shows that even though the wavepacket has spread over
the whole separatrix region, its spread is not uniform. The
wavepacket is strongly localized at the unstable period-$4$
orbit. This strong localization of the wavepacket is also a single top
behavior which may also warrant separate study. 

At $k=3.0$ and $k=6.0$, the initial wavepackets were inside the
chaotic region. However, the saturation of the entanglement production
are different for these two cases. This can be understood as the phase
space of the kicked top is more mixed type for $k=3.0$ than the case
$k=6.0$.  Therefore, the size of the chaotic region is less for
$k=3.0$ than its size corresponding to $k=6.0$. Consequently, the
wavepacket can spread over less of the phase space for $k=3.0$ than
$k=6.0$. In Fig.\ref{fig7}c, we have shown the spreading of the
wavepacket corresponding to this case. At $k=6.0$, since the phase
space is almost fully chaotic, the wavepacket can spread over almost
whole phase space. In Fig.\ref{fig7}d, we have shown reduced Husimi
function corresponding to this strongly chaotic case.

As we know, there exists a universal bound on the entanglement for
chaotic cases and that bound is given by, for the von Neumann entropy,
$(S_V)_{\mbox{sat}} = \ln \gamma N$ where $\gamma =
1/\sqrt{\mbox{e}}\simeq 0.6$. Now a
natural question is whether there exists any such bound on
entanglement of the form $\ln \gamma N^{\prime}$, for the nonchaotic
cases like $k=1.0$ and $k=2.0$. If there exists really such an
entanglement bound, then what is the $N^{\prime}$ in terms of $N$?  We
conjecture that $N^{\prime}$ is actually the effective dimension of the
Hilbert space corresponding to those parameter values, i.e.,
$N^{\prime} = N_{\mbox{eff}} = \Delta N_{\mbox{eff}} N$. Since we know
the evolution of $S_V$ and of $\Delta N_{\mbox{eff}}$, we can
determine the time evolution of that factor $\gamma$ from the relation
\begin{equation}
\gamma = \frac{\exp(S_V)}{N^{\prime}} = \frac{1}{N} \left[ 
\frac{\exp(S_V)}{\Delta N_{\mbox{eff}}} \right].
\label{gamma}
\end{equation}   
\par
\begin{figure}
\centerline{\includegraphics[height=3.5in]{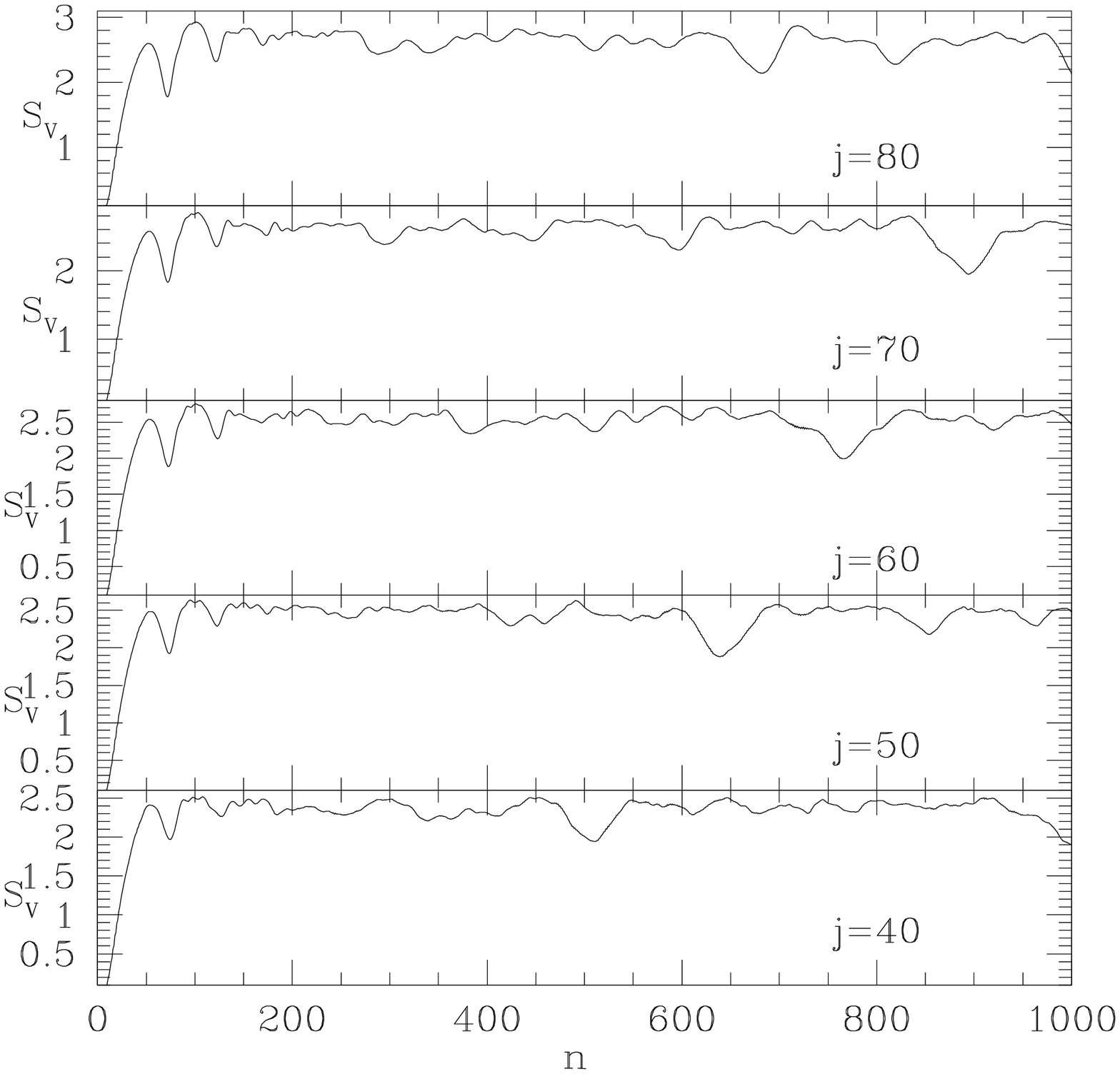}}
\caption{\label{fig8}Evolution of the von Neumann entropy, corresponding to
the parameter value $k=1.0$, are presented for different Hilbert space
dimensions $( N = 2 j + 1 )$.}
\end{figure}
\par
\begin{figure}
\centerline{\includegraphics[height=3.5in]{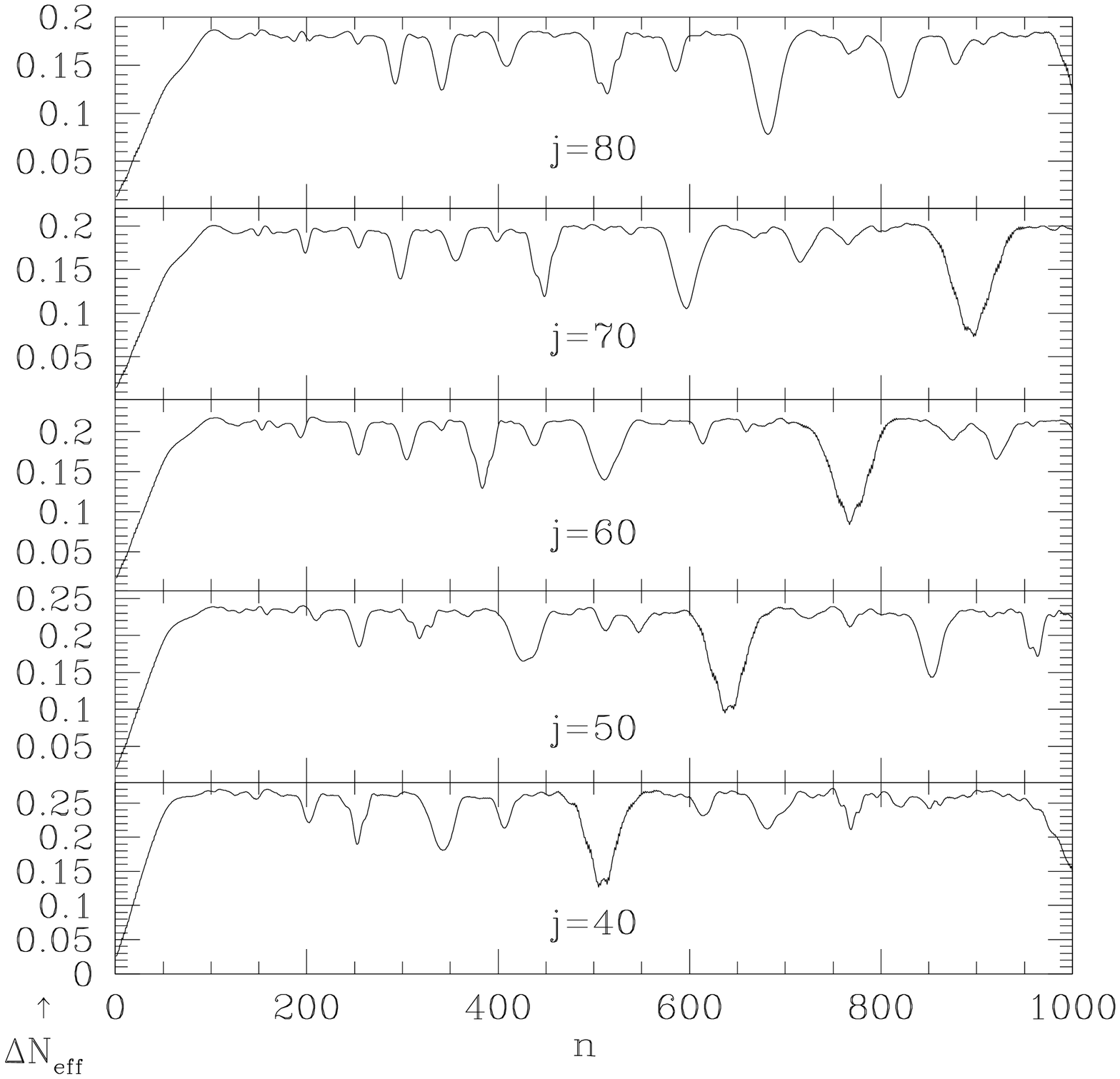}}
\caption{\label{fig9}Evolution of $\Delta N_{\mbox{eff}}$, corresponding to
$k=1.0$, are presented for different Hilbert space dimensions.}
\end{figure}
\par
\begin{figure}
\centerline{\includegraphics[height=3.5in]{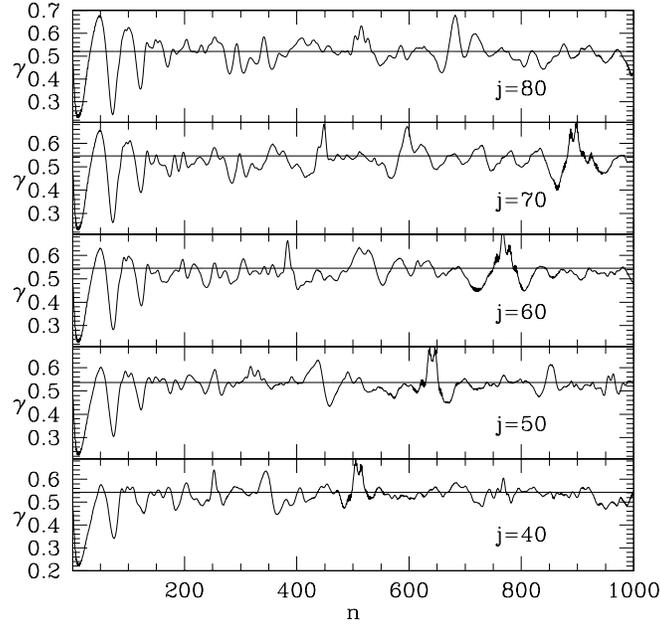}}
\caption{\label{fig10}Evolution of the factor $\gamma$ are presented for
different Hilbert space dimensions. This factor has been calculated numerically
using Eq.(\ref{gamma}). Here $k=1.0$.}
\end{figure}
\par

\begin{figure}
\centerline{\includegraphics[height=3.5in]{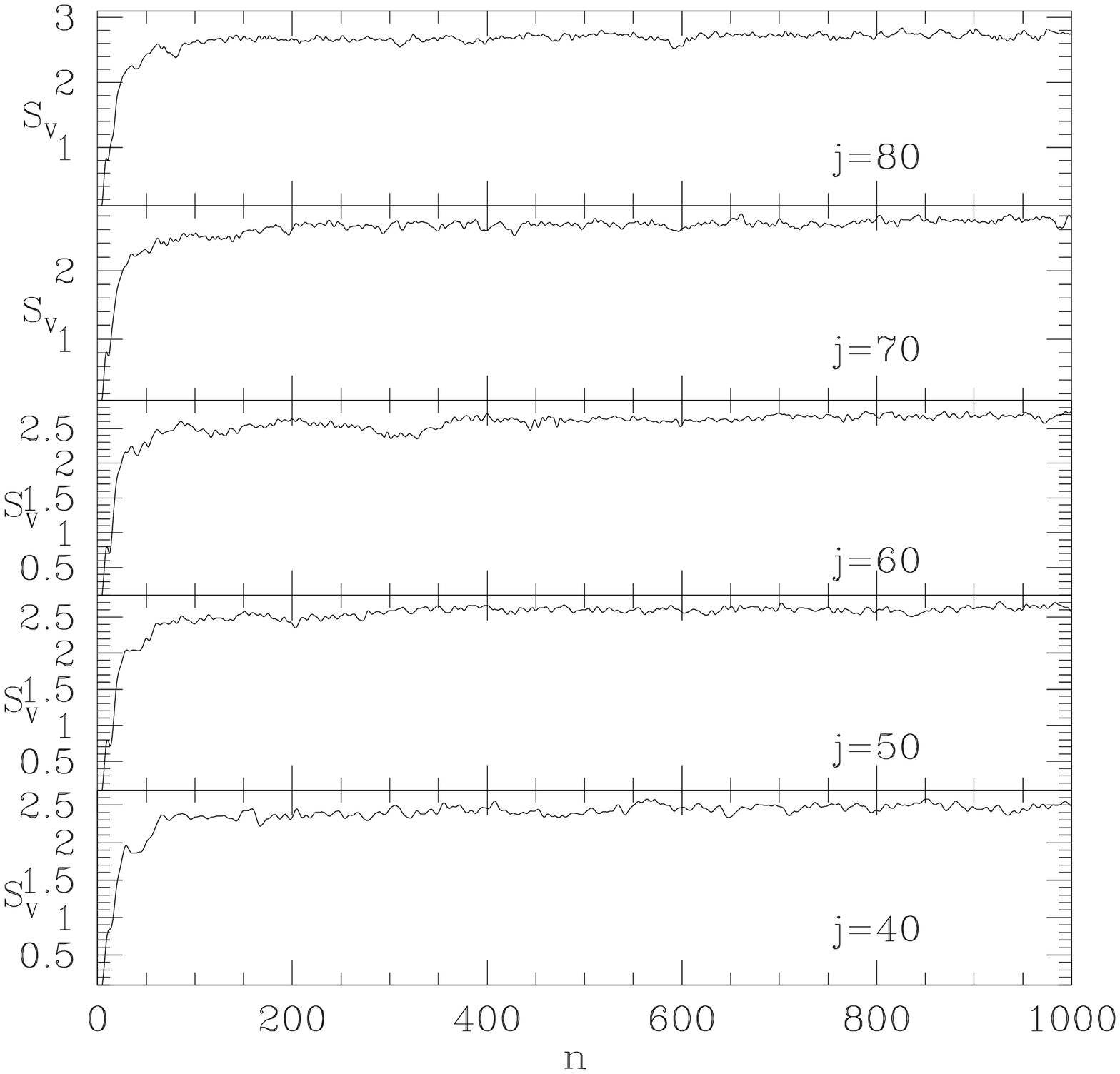}}
\caption{\label{fig11}Evolution of the von Neumann entropy, corresponding to
the parameter value $k=2.0$, are presented for different Hilbert space
dimensions.}
\end{figure}

\begin{figure}
\centerline{\includegraphics[height=3.5in]{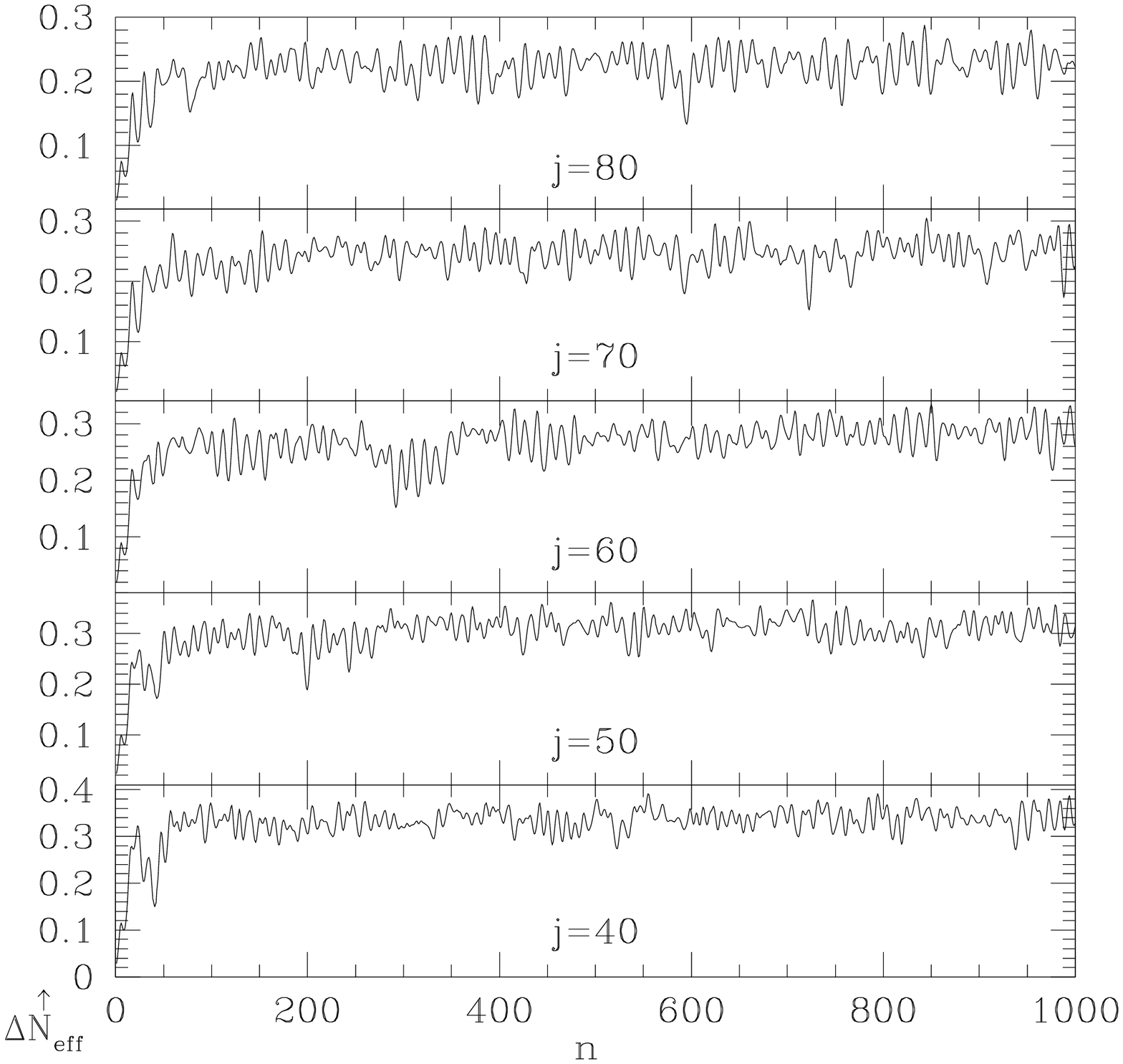}}
\caption{\label{fig12}Evolution of $\Delta N_{\mbox{eff}}$, corresponding to
$k=2.0$, are presented for different Hilbert space dimensions.}
\end{figure}

\begin{figure}
\centerline{\includegraphics[height=3.5in]{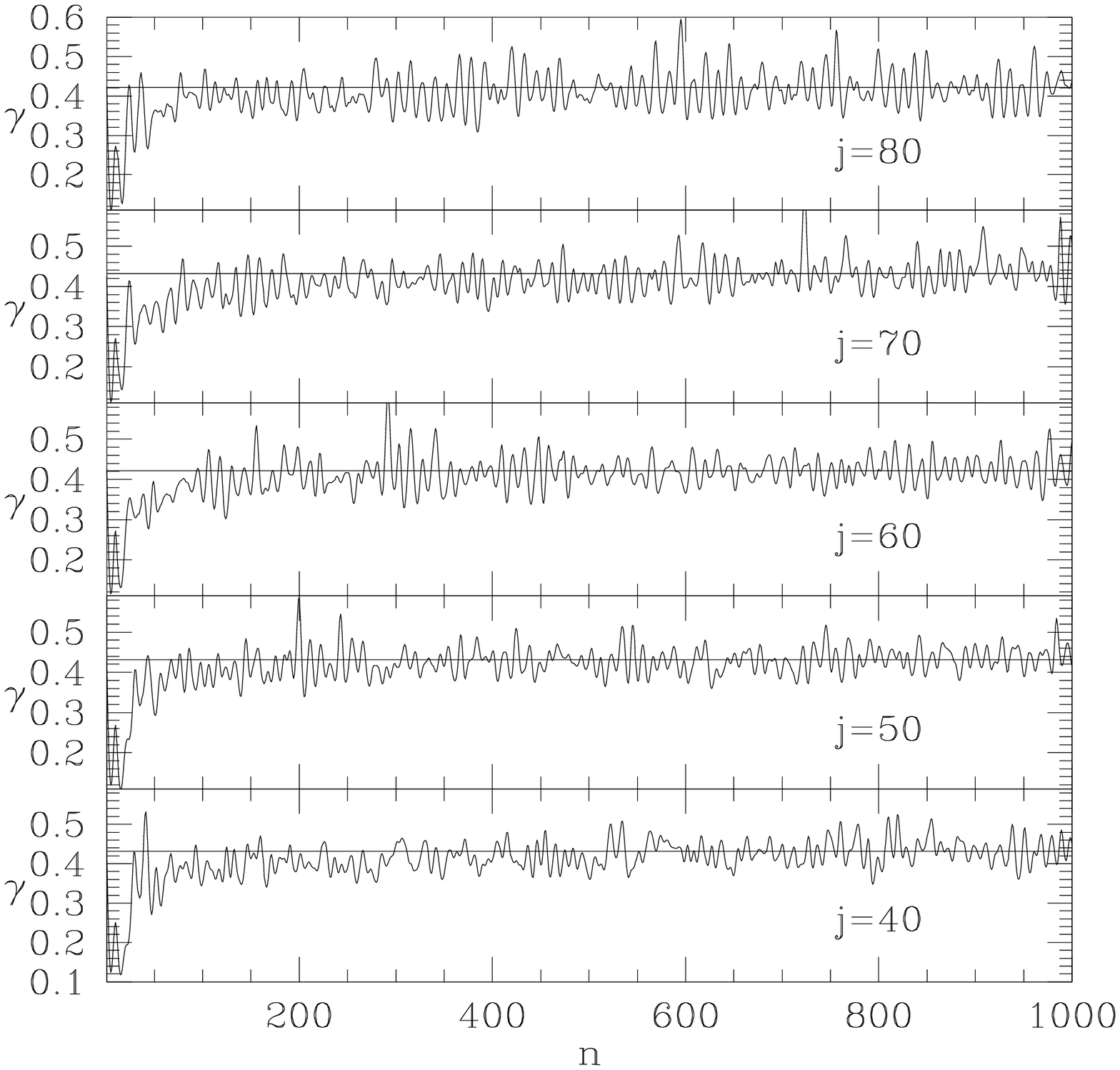}}
\caption{\label{fig13}Evolution of the factor $\gamma$ are presented for
different Hilbert space dimensions. This factor has been calculated numerically
using Eq.(\ref{gamma}). Here $k=2.0$.}
\end{figure}

In Fig.\ref{fig8} and Fig.\ref{fig9}, we have shown the evolution of
$S_V$ and $\Delta N_{\mbox{eff}}$ corresponding to $k=1.0$ for
different Hilbert space dimensions. Using the above relation, we
determine the evolution of $\gamma$ for this $k$-value and that is
presented in Fig.\ref{fig10}. Initially there were some oscillations,
later it fluctuates approximately around $\gamma \simeq 0.52-0.54$ for
different Hilbert space dimensions. The solid line is showing the
average value of $\gamma$ at the saturation region.  Figs.\ref{fig11}
and \ref{fig12} are similarly showing the evolution of $S_V$ and of
$\Delta N_{\mbox{eff}}$ at $k=2.0$ corresponding to different Hilbert
space dimensions. In Fig.\ref{fig13}, we have shown the evolution of
$\gamma$ for this case. This figure is showing that at the saturation
$\gamma \simeq 0.40-0.43$ for different Hilbert space dimensions. At
the saturation, the factor $\gamma$ is different for $k=1.0$ and
$k=2.0$. This is essentially due to the fact that at $k=1.0$ and
$k=2.0$, two different kind of dynamics are responsible for the
spreading of the wavepacket on phase space. At $k=1.0$, the wavepacket
has spread over the regular elliptic orbits, whereas at $k=2.0$ the
wavepacket has spread over a thin stochastic layer present at the
separatrix. Even though we may not expect any universality in the case
of integrable or near-integrable cases, we have found that for a given
coupling strength and for a given classical dynamical behavior, the
factor $\gamma$ is more or less same for different Hilbert space
dimensions.
\par
\par
\par
\par

\subsubsection{Coupling $\epsilon = 10^{-3}$}

\begin{figure}
\centerline{\includegraphics[height=2in]{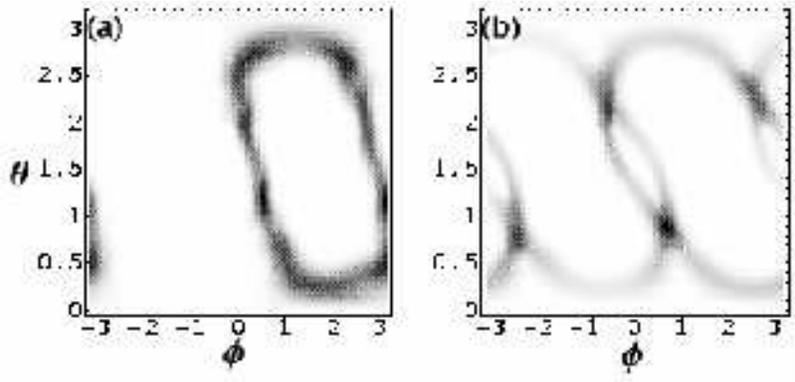}}
\caption{\label{fig14}Reduced Husimi functions of the time evolving wavepacket
are presented corresponding to the time $n=384$ at which the entanglement
production gets saturated. (a) $k=1.0$. The wavepacket is spread over the
elliptic orbits, but the spreading is not uniform. (b) $k=2.0$. The wavepacket
is spread over the separatrix and shows strong localization on the
unstable period-$4$ orbit. Here $\epsilon = 10^{-3}$.}
\end{figure}

Entanglement production corresponding to this coupling strength has
been presented in Fig.\ref{fig7}(b). For the nonchaotic cases ($k=1.0$
and $k=2.0$), the saturation value of the entanglement production is
less than the entanglement saturation observed in the stronger
coupling case ($\epsilon= 10^{-2}$). For weaker coupling, the
influence of one subsystem on the other subsystem becomes less, and
the individual subsystems behave more like isolated quantum
systems. Consequently, pure quantum effects play dominant role in the
evolution of the wavepacket. In Fig.\ref{fig14}, we have shown reduced
Husimi function for $k = 1.0$ and $k = 2.0$ at the time $n = 384$ when
the entanglement production saturated. For $k = 1.0$, the reduced
Husimi function is showing that the wavepacket has spread over the
elliptic orbits, but not uniformly.
\par
\par 

Now for $k = 2.0$, at the entanglement saturation, the wavepacket has
spread as usual over the whole separatrix region. Moreover, it also
shows localization at the same unstable period-$4$ orbit. However, the
difference is that the wavepacket is now more localized at a
particular periodic point of that period-$4$ orbit which was very
close to the initial wavepacket.  As we have seen in Fig. \ref{fig4}b,
within our observational time ($n=1000$), $\Delta N_{\mbox{eff}}$ has
not reached any saturation value for the mixed and as well as for the
chaotic cases. Moreover, for the strong chaos case, $k = 6.0$, the
$\Delta N_{\mbox{eff}}$ was well short of unity even after the
observational time and consequently the wavepacket has not got
access over whole Hilbert space within this time of
observation. Therefore, the entanglement production is well short of
the known statistical bound $\ln(N)-\frac{1}{2}$.

\subsubsection{Coupling $\epsilon = 10^{-4}$}

The entanglement production for this very weak coupling regime has
been presented in Fig.\ref{fig7}(c). The entanglement production for
this weak coupling has recently been explained by perturbation theory
\cite{tanaka}.  However, the formula for the entanglement production
presented in that work is not valid for arbitrarily long times. In the
next section we have presented an approximate formula for the
entanglement production in coupled strongly chaotic systems. This
formula explains the entanglement production for the case $k =
6.0$. Here we have also observed that entanglement production is much
larger for the nonchaotic cases than the chaotic cases. Rather, we can
say that, for weakly coupled cases, the presence of chaos actually
suppresses entanglement production.

\section{\label{analytical}Entanglement production in coupled strongly chaotic 
system}

Due to the relatively simple form of $S_R$, the linear entropy, it is easier 
to derive an approximate formula for its time evolution. Here we present
an analytical formalism for the time evolution of $S_R$ in coupled strongly 
chaotic systems.

Let us assume, the initial state is a product state, given as $|\psi(0)\rangle
= |\phi_1(0)\rangle \otimes |\phi_2(0)\rangle$, where $|\phi_i(0)\rangle$'s are
the states corresponding to individual subsystems. In general, the time
evolution operator of a coupled system is of the form $U \equiv U_{\epsilon}
U_0 = U_{\epsilon} (U_1 \otimes U_2)$, where $U_{\epsilon}$ is the coupling
time evolution operator and $U_i$'s are the time evolution operators of the
individual subsystems. Furthermore, we have assumed
\begin{equation}
U_{\epsilon} = \exp (-i\epsilon H_{12})
\end{equation}
\noindent where $H_{12} = h^{(1)} \otimes h^{(2)}$, and the $h^{(i)}$ are 
Hermitian local operators. For simplicity, we derive our formalism in the 
eigenbasis of $h^{(i)}$'s, i.e., $h^{(i)} |e_{\alpha}^{(i)}\rangle =
e_{\alpha} |e_{\alpha}^{(i)}\rangle$, where $\{e_{\alpha}^{(i)},
|e_{\alpha}^{(i)} \rangle\}$ are the eigenvalues and the corresponding
eigenvectors of $h^{(i)}$. 

The one step operation of $U$ on $|\psi(0)\rangle$ will give,
\begin{equation}
\langle e_{\alpha}^{(1)}, e_{\beta}^{(2)} |\psi(1)\rangle = \exp\left(-i
\epsilon e_{\alpha}^{(1)} e_{\beta}^{(2)}\right)\langle e_{\alpha}^{(1)}
e_{\beta}^{(2)}| \psi_0(1)\rangle,
\end{equation}
\noindent where $|\psi(1)\rangle$ is the time evolving state of the full
coupled system at time $n=1$ and $|\psi_0(1)\rangle$ is the same for the
uncoupled system. From the above expression, one can get the RDM corresponding
to one subsystem by tracing over the other subsystem. The RDM corresponding to
first subsystem is given by,
\begin{eqnarray}
[\rho_1(1)]_{\alpha\beta} &=& \left\langle e_{\alpha}^{(1)}|\rho_1(1)|
e_{\beta}^{(1)}\right\rangle = \sum_{\gamma} \left\langle e_{\alpha}^{(1)},
e_{\gamma}^{(2)}|\psi(1) \right\rangle \left\langle\psi(1)|e_{\beta}^{(1)},
e_{\gamma}^{(2)}\right\rangle\nonumber\\
&=& \sum_{\gamma} \exp\left[-i\epsilon\left(e_{\alpha}^{(1)} - e_{\beta}^{(1)}
\right) e_{\gamma}^{(2)}\right] \left\langle e_{\alpha}^{(1)},e_{\gamma}^{(2)}|
\psi_0(1)\right\rangle \left\langle\psi_0(1)|e_{\beta}^{(1)},e_{\gamma}^{(2)}
\right\rangle.
\end{eqnarray}
\noindent Here we now assume that,  $|\psi_0(1)\rangle$
is a random vector. Consequently we can further assume that the components
of $|\psi_0(1)\rangle$ are uncorrelated to the exponential term coming due to
the coupling. Hence we have,
\begin{eqnarray}
[\rho_1(1)]_{\alpha\beta} &\simeq& \frac{1}{N} \sum_{\gamma} \left\langle
e_{\alpha}^{(1)},e_{\gamma}^{(2)}|\psi_0(1)\right\rangle \left\langle\psi_0(1)|
e_{\beta}^{(1)},e_{\gamma}^{(2)}\right\rangle \sum_{\delta} \exp\left[-i
\epsilon\left(e_{\alpha}^{(1)}-e_{\beta}^{(1)}\right)e_{\gamma}^{(2)}\right]
\nonumber\\
&=& \frac{1}{N} [\rho_{10}(1)]_{\alpha\beta} \sum_{\gamma}\exp\left[-i\epsilon
\left(e_{\alpha}^{(1)}-e_{\beta}^{(1)}\right)e_{\gamma}^{(2)}\right],
\end{eqnarray}
\noindent where $N$ is the Hilbert space dimension of the first subsystem and
$\rho_{10}$ is the density matrix corresponding to the uncoupled top. If we
proceed one more time step, then at the time $n=2$ we have,
\begin{eqnarray}
[\rho_1(2)]_{\alpha\beta} &\simeq& \frac{1}{N} |p(\epsilon)|^2
[\rho_{10}(2)]_{\alpha\beta} \sum_{\gamma} \exp\left[-i\epsilon\left
(e_{\alpha}^{(1)} -e_{\beta}^{(1)}\right)\right]\nonumber\\
\mbox{where}~~~p(\epsilon) &=& \frac{1}{N^2} \sum_{\alpha,\beta} \exp\left(-i
\epsilon e_{\alpha}^{(1)} e_{\beta}^{(2)}\right).
\end{eqnarray}
\noindent If we use the above assumptions upto any arbitrary time $n$, we
obtain
\begin{equation}
[\rho_1(n)]_{\alpha\beta} = \frac{1}{N} |p(\epsilon)|^{2(n-1)}
[\rho_{10}(n)]_{\alpha\beta}\sum_{\gamma}\exp\left[-i\epsilon\left(
e_{\alpha}^{(1)}-e_{\beta}^{(1)}\right)e_{\gamma}^{(2)}\right].
\end{equation}
\noindent From the above expression, it is straightforward to calculate Linear
entropy and that is given as,
\begin{equation}
S_R(n) \simeq 1-\frac{1}{N^4}|p(\epsilon)|^{4(n-1)} \sum_{\alpha,\beta}
\sum_{\gamma,\delta}\exp\left[-i\epsilon\left(e_{\alpha}^{(1)}-e_{\beta}^{(1)}
\right)\left(e_{\gamma}^{(2)}-e_{\delta}^{(2)}\right)\right].
\end{equation}

\noindent This is a general result, applicable to any coupled
strongly chaotic systems of the form $U_{\epsilon} ( U_1 \otimes U_2)$. 
Moreover, this result is valid for long time. 

For the coupled kicked tops $H_{12} = J_{z_1} \otimes J_{z_2}/j$.
Therefore, for this particular system, the above formula would become,
\begin{eqnarray}
S_R(n) &\simeq& 1 - \frac{1}{N^4} p(\epsilon)^{4(n-1)}
\sum_{m_1, n_1 =-j}^{+j} \sum_{m_2, n_2 = -j}^{+j} \exp\left[ - i
\frac{\epsilon}{j} (m_1 - n_1)(m_2 - n_2) \right]\nonumber\\
\mbox{where}~~p(\epsilon) &=& \frac{1}{N^2} \sum_{m_1, m_2 = -j}^{+j}
\exp\left(-i\frac{\epsilon}{j} m_1 m_2 \right)~~\mbox{and}~~N = 2 j + 1
\label{anal_for1}
\end{eqnarray}
\noindent In large $j$-limit, we can substitute above sums by approximate 
integrals and then performing those integrals we get (for details, see
Appendix \ref{d}),
\begin{eqnarray}
S_R(n) \simeq 1 - p(\epsilon)^{4(n-1)} \left[ \frac{2}{N}\left\{ 1 +
\frac{\mbox{Si}(2N\epsilon)}{\epsilon}\right\}\right. &-& \left.
\left(\frac{1}{N\epsilon} \right)^2 \left\{1 - \cos(2N\epsilon) + 
\mbox{Ci}(2N\epsilon) - \ln(2N\epsilon) - \gamma\right\}\right]\nonumber\\
\mbox{where}~~~p(\epsilon) &\simeq& \frac{2}{N} \left[ 1 +
\frac{1}{\epsilon} \mbox{Si}\left(\frac{N \epsilon}{2}\right)\right].
\label{anal_for2}
\end{eqnarray}
\noindent The functions Si and Ci are the standard {\it Sine-integral} and
{\it Cos-integral} function, while $\gamma = 0.577216 \ldots$ is the
Euler constant. In the above derivation we have not assumed, unlike
the perturbation theory \cite{tanaka}, any particular order of
magnitude of the coupling strength $\epsilon$. Therefore, as we
demonstrate below the above formula is applicable for non-perturbative
coupling strengths as well.
\par
\begin{figure}
\centerline{\includegraphics[height=3.5in]{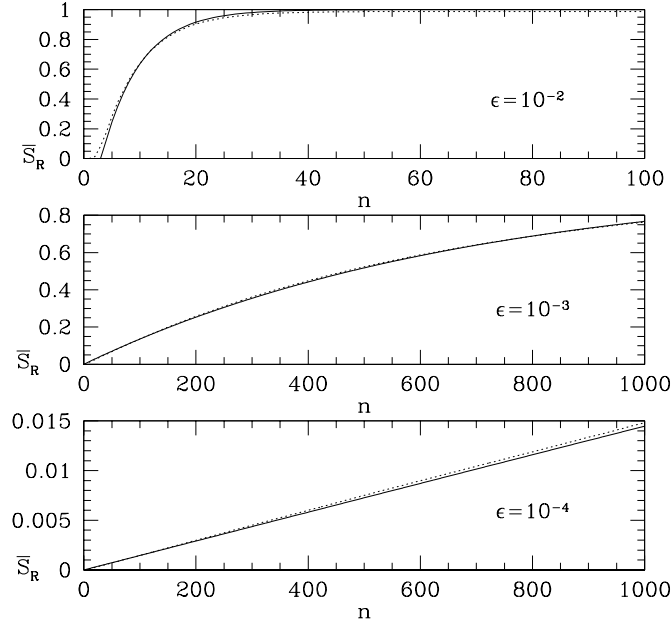}}
\caption{\label{fig15}Evolution of the Linear entropy for the coupled
strongly chaotic system is presented. The dotted line is the numerical results
of the coupled kicked tops system. We choose $k=6.0$ for the first top and
$k=6.1$ for the second top. The solid line is the theoretical estimation,
given by Eq.(\ref{anal_for2}).}
\end{figure}
\par

In Fig.\ref{fig15}, we have shown the numerical result of the Linear
entropy ($S_R$) production in the coupled tops where the individual
tops are strongly chaotic. Here we have considered many initial
coherent states at different parts of the phase space and presented
the Linear entropy production, averaged over all these initial states,
with time. In all our previous calculations we only considered the
entanglement production on coupling identical tops, therefore,
permutation symmetry was present. As in the above derivation, we have
not assumed any special symmetry property, we break permutation
symmetry by taking slightly non-identical tops with $k = 6.0$ for the
first top and $k = 6.1$ for the second top. 

Fig.\ref{fig15} demonstrates that our theoretical estimation, denoted
by the solid curve, is not only valid for weak coupling case like
$\epsilon = 10^{-4}$ but it also valid for sufficiently strong
coupling cases like $\epsilon = 10^{-2}$. Moreover, this formula is
applicable for very long times. If we consider weak coupling
approximation, i.e., $j \epsilon \ll 1$, then the above formula will
become approximately,
\begin{equation}
S_R(n) \simeq \frac{2 \epsilon^2 j^2}{9} ( n - 1) + {\cal O}(\epsilon^3 j^3).
\end{equation}
\noindent Therefore, at this weak coupling approximation, the entanglement
production rate is $2 \epsilon^2 j^2 / 9$, which has been calculated in a 
recent publication \cite{tanaka} by very different means.

\section{\label{last}Summary}

In this paper, our major goal was to study entanglement production in
coupled kicked tops. Single kicked top is a well studied model of both
classical and quantum chaotic system. The classical map corresponding
to coupled kicked tops was presented in a previous publication, but
was unfortunately incorrect. Hence, we have presented the correct
classical map corresponding to the coupled kicked tops which is
canonical. In the quantum case, we have studied the
reduced Husimi function to visualize the behavior of the wavepacket
of a coupled system in any one of it subspaces. We have also studied a
phase space based measure of complexity of the time evolving state (
denoted by $\Delta N_{\mbox{eff}}$ ), which quantify the fraction of
the total number of the Planck cells occupied by the Husimi function
of a given state. As we have already mentioned that, for kicked top,
this quantity is also approximately equal to the fraction of the
Hilbert space occupied by a given state. We have studied this quantity
for both single and coupled tops. It has been observed that, for the
single top, the time evolving state can occupy maximum, in average,
{\it half} of the total number of the Planck cell, i.e., $\Delta
N_{\mbox{eff}} = 0.5$, and this happened for the strongly chaotic
cases. 

For nonchaotic and mixed cases, the time evolving state occupies even
less number of Planck cells and it is reflected in smaller values of
$\Delta N_{\mbox{eff}}$. Following a recent result, using RMT, we have
explained the fact that $\Delta N_{\mbox{eff}} = 0.5$ for the time
evolving state corresponding to strongly chaotic single top. However,
when a strongly chaotic top is strongly coupled to another such top,
$\Delta N_{\mbox{eff}}$ corresponding to any subsystem reaches very
close to {\it unity}. We have again explained this by means of RMT
calculations. 

Then we studied entanglement production in coupled kicked
tops for different underlying classical dynamics of the individual top
and also for different coupling strengths. We find, in general,
entanglement production is higher for stronger chaotic
cases. Moreover, coupling strength between two tops is also an
important parameter for the entanglement production. For example, when
the coupling strength between two tops is very weak, we find that
entanglement production is higher for sufficiently long time
corresponding to nonchaotic cases. Finally, we have derived an
approximate formula, based on the ideas of RMT, for the entanglement
production in coupled strongly chaotic system. This formula is
applicable, unlike perturbation theory, to large coupling
strengths and is valid for sufficiently long times.

\appendix

\section{\label{a} Derivation of Eq.(\ref{Jx1})}

Let us define ladder operators,
\begin{eqnarray}
J_{1\pm} \equiv J_{x_1} \pm J_{y_1}~&;&~~~J_{1+} = J_{1-}\nonumber\\
J_{1+} |m_1\rangle = C_{m_1} |m_1+1\rangle &\mbox{and}& J_{1-} |m_1\rangle
= D_{m_1} |m_1-1\rangle
\end{eqnarray}
\noindent where $C_{m_1}$ and $D_{m_1}$ are known functions of $j$ and $m_1$
and ${|m_1\rangle}$ are the standard angular momentum basis states. We can
write $J_{x_1} = (J_{1+} + J_{1-})/2$. Therefore,
\begin{equation}
J_{x_1}^{\prime} \otimes I_2 = \frac{1}{2} U_{T}^{\dagger} (J_{1+} \otimes I_2)
U_T + \frac{1}{2} U_{T}^{\dagger} (J_{1-} \otimes I_2) U_T,
\label{A1_1}
\end{equation}
\noindent where the terms present at the right hand side are the Hermitian
conjugate of each other. Therefore, it is sufficient to determine only
one term. Here we will calculate the first term explicitly. We have
\begin{equation}
U_{T}^{\dagger} ( J_{1+} \otimes I_2 ) = ( U_1 \otimes U_2 )^{\dagger}
U_{12}^{\epsilon^{\dagger}} ( J_{1+} \otimes I_2 ) U_{12}^{\epsilon}
( U_1 \otimes U_2 ).
\end{equation}
\noindent In $| m_1, m_2 \rangle$ basis, $U_{12}^{\epsilon^{\dagger}}
( J_{1+} \otimes I_2 ) U_{12}^{\epsilon}$ is,
\begin{eqnarray}
&&\langle m_1, m_2 | U_{12}^{\epsilon^{\dagger}} ( J_{1+} \otimes I_2 )
U_{12}^{\epsilon} | n_1, n_2 \rangle \nonumber\\
&=& \exp\left[ i \frac{\epsilon}{j} (m_1 - n_1) m_2 \right] \langle
m_1 | J_{1+} | n_1 \rangle \delta_{m_2 n_2}\nonumber\\
&=& \exp\left[ i \frac{\epsilon}{j} (m_1 - n_1) m_2 \right] C_{n_1}
\delta_{m_1, n_1 + 1} \delta_{m_2 n_2} \nonumber\\
&=& \exp\left[ i \frac{\epsilon}{j} m_2 \right] C_{n_1} \delta_{m_1, n_1 + 1}
\delta_{m_2 n_2}.
\end{eqnarray}
\noindent The above expression can also be written as,
\begin{eqnarray}
\langle m_1, m_2 | U_{12}^{\epsilon^{\dagger}} ( J_{1+} \otimes I_2 )
U_{12}^{\epsilon} | n_1, n_2 \rangle &=& \langle m_1, m_2 | J_{1+} \otimes
\exp\left(i \frac{\epsilon}{j} J_{z_2} \right) | n_1, n_2 \rangle\nonumber\\
\Rightarrow U_{12}^{\epsilon^{\dagger}} ( J_{1+} \otimes I_2 ) U_{12}^{\epsilon}
&=& J_{1+} \otimes \exp\left( i \frac{\epsilon}{j} J_{z_2} \right).
\end{eqnarray}
\noindent Therefore,
\begin{eqnarray}
U_{T}^{\dagger} ( J_{1+} \otimes I_2 ) U_T &=& ( U_1 \otimes U_2 )^{\dagger}
\left[ J_{1+} \otimes \exp\left(i \frac{\epsilon}{j} J_{z_2} \right)\right]
( U_1 \otimes U_2 )\nonumber\\
&=& ( U_{1}^{\dagger} J_{1+} U_1 ) \otimes \left[ U_{2}^{\dagger}
\exp\left(i \frac{\epsilon}{j} J_{z_2} \right) U_2 \right]
\label{A1_2}
\end{eqnarray}
\noindent Now,
\begin{eqnarray}
U_{1}^{\dagger} J_{1+} U_1 &=& U_{1}^{f^{\dagger}} U_{1}^{k^{\dagger}} J_{1+}
U_{1}^{k} U_{1}^{f} \nonumber\\
&=& U_{1}^{f^{\dagger}} J_{1+}^{\prime\prime} U_{1}^{f},~~\mbox{where}~~
J_{1+}^{\prime\prime} \equiv U_{1}^{k^{\dagger}} J_{1+} U_{1}^{k}.
\end{eqnarray}
\noindent In $\left\{ | m_1 \rangle \right\}$ basis, $J_{1+}^{\prime\prime}$ 
can be written as,
\begin{eqnarray}
\langle m_1 | J_{1+}^{\prime\prime} | n_1 \rangle &=& \langle m_1 |
U_{1}^{k^{\dagger}} J_{1+} U_{1}^{k} | n_1 \rangle, \nonumber \\
&=& \exp\left[ i \frac{k}{2j} \left( m_{1}^{2} - n_{1}^{2} \right) \right]
\langle m_1 | J_{1+} | n_1 \rangle,\nonumber\\
&=& \exp\left[ i \frac{k}{2j} \left( m_{1}^{2} - n_{1}^{2} \right) \right]
C_{n_1} \delta_{m_1, n_1 + 1}, \nonumber\\
&=& \exp\left[ i \frac{k}{j} \left( n_1 + \frac{1}{2} \right) \right] C_{n_1}
\delta_{m_1, n_1 + 1} \nonumber,\\
&=& \langle m_1 | J_{1+} \exp\left[ i \frac{k}{j} \left( J_{z_1} + \frac{1}{2}
\right) \right] | n_1 \rangle\nonumber\\
\Rightarrow J_{1+}^{\prime\prime} &=& J_{1+} \exp\left[ i \frac{k}{j} \left(
J_{z_1} + \frac{1}{2} \right) \right]
\end{eqnarray}
\noindent Therefore,
\begin{equation}
U_{1}^{\dagger} J_{1+} U_1 = U_{1}^{f^{\dagger}} J_{1+} \exp\left[ i
\frac{k}{j} \left( J_{z_1} + \frac{1}{2} \right) \right] U_{1}^{f}.
\end{equation}
\noindent The operator $U_{1}^{f}$ is the rotation operator about $y-$axis
with angle $\pi/2$, therefore $U_{1}^{f^{\dagger}} ( J_{x_1}, J_{y_1}, J_{z_1} )
U_{1}^{f} = ( J_{z_1}, J_{y_1}, - J_{x_1} )$. Hence we have
\begin{equation}
U_{1}^{\dagger} J_{1+} U_1 = ( J_{z_1} + i J_{y_1} ) \exp\left[ i \frac{k}{j}
\left( - J_{x_1} + \frac{1}{2} \right)\right].
\end{equation}
\noindent Now we will calculate the other term of Eq.(\ref{A1_2}), i.e.,
\begin{eqnarray}
U_{2}^{\dagger} \exp\left( i \frac{\epsilon}{j} J_{z_2} \right) U_2
&=& U_{2}^{f^{\dagger}} U_{2}^{k^{\dagger}} \exp\left( i \frac{\epsilon}{j}
J_{z_2} \right) U_{2}^{k} U_{2}^{f}\nonumber\\
&=& U_{2}^{f^{\dagger}} \exp\left( i \frac{\epsilon}{j} J_{z_2} \right)
U_{2}^{f}~~[~\mbox{since}~ \left[ U_{2}^{k}, J_{z_2} \right] = 0]\nonumber\\
&=& \exp\left( - i \frac{\epsilon}{j} J_{x_2} \right) ~~~\left[~\mbox{since}~   
U_{2}^{f} ~\mbox{is} \: \mbox{rotation} \: \mbox{matrix} \right].
\end{eqnarray}
\noindent Substituting all the above results in Eq. (\ref{A1_2}), we get,
\begin{equation}
U_{T}^{\dagger} ( J_{1+} \otimes I_2 ) U_T = ( J_{z_1} + i J_{y_1} )
\exp\left[ i \frac{k}{j} \left( - J_{x_1} + \frac{1}{2} \right) \right] \otimes
\exp\left( - i \frac{\epsilon}{j} J_{x_2} \right).
\end{equation}
\noindent By taking Hermitian conjugate of the above expression, we determine
\begin{equation}
U_{T}^{\dagger} ( J_{1-} \otimes I_2 ) U_T = \exp\left[-i\frac{k}{j}
\left( - J_{x_1} + \frac{1}{2} \right) \right] ( J_{z_1} - i J_{y_1} ) \otimes
\exp\left( i \frac{\epsilon}{j} J_{x_2} \right).
\end{equation}
\noindent Substituting, last two expressions in Eq.(\ref{A1_1}), we will get
Eq.(\ref{Jx1}).

\section{\label{c}Calculation of the integral present in Eq. (\ref{integral})
and Eq.(\ref{M2rho1H})}

We know $\langle m | z \rangle = \langle m | \theta, \phi \rangle$ and using 
Eq.(\ref{costate}), the above mentioned integral becomes
\begin{eqnarray}
&&\frac{2 j + 1}{4 \pi} \sqrt{
\left( \begin{array}{c} 2 j \\ j - i \end{array} \right)
\left( \begin{array}{c} 2 j \\ j - k \end{array} \right)
\left( \begin{array}{c} 2 j \\ j - l \end{array} \right)
\left( \begin{array}{c} 2 j \\ j - m \end{array} \right)}
\int_{\theta = 0}^{\pi} \int_{\phi = - \pi}^{\pi} \left( 1 + \tan^2 
\frac{\theta}{2} \right)^{-4 j} \left( \tan \frac{\theta}{2} 
\right)^{4 j - i - k - l -m}\nonumber\\ 
&&\exp[-i\phi \{ ( i + l ) - ( k + m ) \} ] \sin\theta d\theta d\phi.
\end{eqnarray}
\noindent After performing the $\phi$-integral, we get
\begin{equation}
(2 j + 1) \sqrt{
\left( \begin{array}{c} 2 j \\ j - i \end{array} \right)
\left( \begin{array}{c} 2 j \\ j - k \end{array} \right)
\left( \begin{array}{c} 2 j \\ j - l \end{array} \right)
\left( \begin{array}{c} 2 j \\ j - m \end{array} \right)}
\delta_{i+l,k+m}
\int_{\theta = 0}^{\pi} \left( \cos \frac{\theta}{2} \right)^{4j + 2(i+l) + 1}
\left( \sin\frac{\theta}{2} \right)^{4j - 2(i+l) + 1} d\theta. 
\end{equation}
\noindent Substituting $\eta = \theta/2$, we get
\begin{equation} 
2 ( 2 j + 1) \sqrt{
\left( \begin{array}{c} 2 j \\ j - i \end{array} \right)
\left( \begin{array}{c} 2 j \\ j - k \end{array} \right)
\left( \begin{array}{c} 2 j \\ j - l \end{array} \right)
\left( \begin{array}{c} 2 j \\ j - m \end{array} \right)}
\delta_{i+l,k+m}
\int_{\eta = 0}^{\frac{\pi}{2}} (\sin\eta)^{4 j - 2 ( i + l ) + 1}
(\cos\eta)^{4 j + 2 ( i + l ) + 1} d\eta.
\end{equation}
\noindent The above integral is a $\beta$-integral, and therefore we get,
\begin{equation}
( 2 j + 1 ) \sqrt{
\left( \begin{array}{c} 2 j \\ j - i \end{array} \right)
\left( \begin{array}{c} 2 j \\ j - k \end{array} \right)
\left( \begin{array}{c} 2 j \\ j - l \end{array} \right)
\left( \begin{array}{c} 2 j \\ j - m \end{array} \right)}
\beta[ \{ (2 j + 1) - (i + l) \}, \{ (2 j + 1) + (i + l) \} ]
\delta_{i+l,k+m}
\end{equation}
\noindent From the relation, $ \beta ( m, n ) = \left[ \Gamma(m) \Gamma(n)
\right] / \Gamma(m+n)$, we get
\begin{equation}
\frac{2 j + 1}{\Gamma(4 j + 2)} \sqrt{
\left( \begin{array}{c} 2 j \\ j - i \end{array} \right)
\left( \begin{array}{c} 2 j \\ j - k \end{array} \right)
\left( \begin{array}{c} 2 j \\ j - l \end{array} \right)
\left( \begin{array}{c} 2 j \\ j - m \end{array} \right)}
\Gamma\{ (2 j + 1) - (i + l)\} \Gamma\{ (2 j + 1) + (i + l) \} 
\delta_{i+l,k+m}.
\end{equation} 
\noindent We know that, for any integer $m$, $\Gamma(m+1) = m !$. Using this
relation the above expression will be equal to Eq.(\ref{sol1}) and
Eq.(\ref{sol2}). 

\section{\label{d} Calculation of Eq.(\ref{anal_for2})}

Let us first calculate the sum present in the expression of $p(\epsilon)$.
That sum can be simplified in the following way.
\begin{eqnarray}
&& \sum_{m_1, m_2 = -j}^{+j} \exp\left( -i \frac{\epsilon}{j} m_1 m_2 \right)
\nonumber\\
&=& \frac{2 N - 1}{N^2} + \frac{1}{N^2} \left[ \sum_{m_1 = 1}^{j}
\sum_{m_2 = 1}^{j} \exp\left( -i \frac{\epsilon}{j} m_1 m_2 \right)
+ \sum_{m_1 = -j}^{-1} \sum_{m_2 = 1}^{j} \exp\left( -i \frac{\epsilon}{j} m_1
m_2 \right)\right. \nonumber\\
&+&  \left. \sum_{m_1 = 1}^{j}
\sum_{m_2 = -j}^{-1} \exp\left( -i \frac{\epsilon}{j} m_1 m_2 \right)
+ \sum_{m_1 = -j}^{-1} \sum_{m_2 = -j}^{-1} \exp\left( -i \frac{\epsilon}{j}
m_1 m_2 \right)\right] \nonumber\\
&=& \frac{2 N - 1}{N^2} + \frac{2}{N^2} \left[\sum_{m_1, m_2 = 1}^{j}
\exp\left( i \frac{\epsilon}{j} m_1 m_2 \right) + \sum_{m_1, m_2 = 1}^{j}
\exp\left( - i \frac{\epsilon}{j} m_1 m_2 \right)\right]\nonumber\\
&=& \frac{2 N - 1}{N^2} + \frac{4}{N^2} \mbox{Re} \sum_{m_1, m_2 = 1}^{j}
\exp\left( i \frac{\epsilon}{j} m_1 m_2 \right),
\label{p_sum}
\end{eqnarray}
\noindent where `Re' denoting the real part. Now we define $x \equiv m_1/j$,
$y \equiv m_2/j$ and $\delta \equiv 1/j$, where $\delta \rightarrow 0$ in large
$j$ limit. We can convert the above sum into an integral in the large
$j$-limit as,
\begin{eqnarray}
&& j^2 \lim_{\delta \rightarrow 0} \int_{x = \delta}^{1} \int_{y = \delta}^{1}
d x d y \cos (j \epsilon x y)\nonumber\\
&=& j^2 \lim_{\delta\rightarrow 0} \int_{x = \delta}^{1}
\frac{\sin(j\epsilon x)}{j \epsilon x} d x \nonumber\\
&=& \frac{j}{\epsilon} \mbox{Si}(j \epsilon)
\end{eqnarray}
\noindent In the large $j$-limit, $N = 2 j + 1 \simeq 2 j$, therefore
\begin{equation}
p(\epsilon) \simeq \frac{2 N - 1}{N^2} + \frac{2}{N \epsilon} \mbox{Si}
\left(\frac{N \epsilon}{2}\right).
\end{equation}
\noindent If we neglect $N^{-2}$ term, then we get,
\begin{equation}
p(\epsilon) \simeq \frac{2}{N} \left[ 1 + \frac{\mbox{Si}\left(
\frac{N \epsilon}{2}\right)}{\epsilon}\right].
\end{equation}

Let us now calculate the bigger sum [see Eq. (\ref{anal_for1})]. If we define
$l_1 \equiv m_1 - n_1$ and $l_2 \equiv m_2 - n_2$, then this sum will become,
\begin{eqnarray}
&&\sum_{l_1, l_2 = -M}^{+M} ( N - |l_1| ) ( N - |l_2| )
\exp\left(-i \frac{\epsilon}{j} l_1 l_2 \right)~;~~M = 2 j = N - 1\nonumber\\
&=& 2 N \sum_{l_1 = -M}^{+ M} ( N - |l_1| ) +
\sum_{\begin{array}{c} l_1 = -M \\ l_1 \neq 0 \end{array}}^{+ M}
\sum_{\begin{array}{c} l_2 = -M \\ l_2 \neq 0 \end{array}}^{+ M}
( N - |l_1| ) ( N - |l_2| ) \exp\left( -i \frac{\epsilon}{j} l_1 l_2 \right)
\nonumber\\
&=& 4 N^2 M + 4 \mbox{Re} \sum_{l_1, l_2 = 1}^{M} ( N - l_1 ) ( N - l_2 )
\exp\left( i \frac{\epsilon}{j} l_1 l_2 \right)\nonumber\\
&=& 4 N^2 M + 4 N^2 \mbox{Re} \sum_{l_1, l_2 = 1}^{M} \exp\left(i
\frac{\epsilon}{j} l_1 l_2 \right) - 8 N \mbox{Re} \sum_{l_1, l_2 = 1}^{M}
l_1 \exp\left( i \frac{\epsilon}{j} l_1 l_2 \right) \nonumber\\
&+& 4 \mbox{Re} \sum_{l_1, l_2 = 1}^{M} l_1 l_2 \exp\left( i \frac{\epsilon}{j}
l_1 l_2 \right)
\label{big_sum}
\end{eqnarray}
\noindent We can write the first sum of the above expression as,
\begin{equation}
\sum_{l_1, l_2 = 1}^{M} \exp\left( i \frac{\epsilon}{j} l_1 l_2 \right)
= \sum_{l_1, l_2 = 1}^{M} \exp\left(i \frac{2 \epsilon}{M} l_1 l_2 \right).
\end{equation}
\noindent This sum is similar to the sum which we have calculated to derive
$p(\epsilon)$, see Eq. (\ref{p_sum}). Therefore, using this previous
result, we get the above sum as,
\begin{equation}
\sum_{l_1,l_2 = 1}^{M} \exp\left( i \frac{\epsilon}{j} l_1 l_2 \right)
\simeq \frac{M}{2 \epsilon} \mbox{Si} ( 2 M \epsilon).
\end{equation}
\noindent Now
\begin{eqnarray}
\mbox{Second}~\mbox{sum}~ &=& \mbox{Re} \sum_{l_1, l_2 = 1}^{M} l_1 \exp\left(
i \frac{\epsilon}{j} l_1 l_2 \right)\nonumber\\
&\simeq& M^3 \lim_{\delta\rightarrow 0} \int_{x = \delta}^{1}
\int_{y = \delta}^{1} x \cos( 2 M \epsilon x y ) d x d y \nonumber\\
&\simeq& \frac{M^2}{2 \epsilon} \int_{0}^{1} \sin( 2 M \epsilon x ) d x
\nonumber\\
&\simeq& \frac{M}{4 \epsilon^2} [ 1 - \cos (2 M \epsilon ) ]\\
\mbox{Third}~\mbox{sum}~ &=& \mbox{Re} \sum_{l_1, l_2 = 1}^{M} l_1 l_2
\exp\left( i \frac{\epsilon}{j} l_1 l_2 \right)\nonumber\\
&\simeq& M^4 \lim_{\delta\rightarrow 0} \int_{x = \delta}^{1}
\int_{y = \delta}^{1} d x d y x y \cos (2 M \epsilon x y)\nonumber\\
&\simeq& \frac{M^3}{2 \epsilon} \lim_{\delta\rightarrow 0} \left[
\int_{x = \delta}^{1} \sin (2 M \epsilon x) d x + \int_{x = \delta}^{1}
\frac{\cos (2 M \epsilon x) - 1}{2 M \epsilon x} d x \right] \nonumber\\
&\simeq& \frac{M^2}{4 \epsilon^2} [ 1 - \cos (2 M \epsilon) + \mbox{Ci}
( 2 M \epsilon ) - \ln (2 M \epsilon) - \gamma ].
\end{eqnarray}
\noindent For large $j$-limit, $M \simeq N$ and therefore substituting above 
results in Eq.(\ref{big_sum}), we will arrive at Eq.(\ref{anal_for2}).

\end{document}